\documentclass[12pt]{article}
\usepackage{amsmath,amsthm,amssymb,mathtools}
\usepackage{graphicx,epstopdf,framed,color}
\usepackage{euscript,paralist,wrapfig,}
\usepackage[T1]{fontenc}
\usepackage[utf8]{inputenc}
\usepackage{tikz,float}

\theoremstyle{remark}
\newtheorem{theorem}{Theorem}[section]

\newtheorem{assumption}[theorem]{Assumption}

\newtheorem{proposition}[theorem]{Proposition}

\newtheorem{remark}[theorem]{Remark}

\newtheorem{problem}[theorem]{Problem}

\numberwithin{equation}{section}

\newfont{\roma}{cmr10 scaled 1200}
\renewcommand{\cline}{{\mathbb C}}

\newcommand{\nline}  {{\mathbb N}}
\newcommand{\rline}  {{\mathbb R}}

\newcommand{\tline}  {{\mathbb T}}

\newcommand{\dd}   {{\rm d}\hbox{\hskip 0.5pt}}

\newcommand{\Dscr} {{\cal D}}

\newcommand{\Jscr} {{\cal J}}

\newcommand{\Lscr} {{\cal L}}

\newcommand{\Qscr} {{\cal Q}}

\newcommand{\Sscr} {{\cal S}}

\newcommand{\mm}    {{\hbox{\hskip 0.5pt}}}
\newcommand{\m}     {{\hbox{\hskip 1pt}}}

\newcommand{\bluff} {{\hbox{\raise 15pt \hbox{\mm}}}}
\newcommand{\sbluff}{{\hbox{\raise  7pt \hbox{\mm}}}}

\newcommand{\g}      {{\gamma}}
\newcommand{\FORALL} {{\hbox{$\hskip 11mm \forall \;$}}}

\renewcommand{\Re}   {{\rm Re\,}}

\newcommand{\bbm}[1]{\left[\begin{matrix} #1 \end{matrix}\right]}
\newcommand{\sbm}[1]{\left[\begin{smallmatrix} #1
   \end{smallmatrix}\right]}

\begin{document}
%%%%%%%%%%**********%%%%%%%%%%**********%%%%%%%%%%**********%%%%%%%%%%
\renewcommand{\thefootnote}{\fnsymbol{footnote}}
\renewcommand{\thefootnote}{\fnsymbol{footnote}}
\newcommand{\footremember}[2]{%
   \footnote{#2}
    \newcounter{#1}
    \setcounter{#1}{\value{footnote}}%
}
\newcommand{\footrecall}[1]{%
    \footnotemark[\value{#1}]%
}
\makeatletter
\def\blfootnote{\gdef\@thefnmark{}\@footnotetext}
\makeatother

\begin{center}
{\Large \bf LQR based $\omega-$stabilization of a heat equation \\[1ex]
 with memory}\\[2ex]
Bhargav Pavan Kumar Sistla, Wasim Akram, Debanjana Mitra, Vivek Natarajan \vspace{2mm}
\blfootnote{\!\!\!\!\!\!B. P. K. Sistla is supported by the Prime Minister’s Research Fellowship grant RSPMRF0262.}
\blfootnote{\!\!\!\!\!\!B. P. K. Sistla (bhargav@sc.iitb.ac.in) and V. Natarajan (vivek.natarajan@iitb.ac.in) are with the Centre for Systems and Control and D. Mitra (deban@math.iitb.ac.in) is with the Department of Mathematics, Indian Institute of Technology Bombay, Mumbai, India, 400076, Ph:+912225765385.}
\blfootnote{\!\!\!\!\!\!W. Akram (wakram2k11@gmail.com) is with the Department of Mathematics, Indian Institute of Technology Roorkee, Roorkee, India, 247667.}
\end{center}

\begin{center}
  {\bf Abstract}\vspace{2mm}

\parbox{5.5in}{{\noindent
We consider a heat equation with memory which is defined on a bounded domain in $\rline^d$ and is driven by $m$ control inputs acting on the interior of the domain. Our objective is to numerically construct a state feedback controller for this equation such that, for each initial state, the solution of  the closed-loop system decays exponentially to zero with a decay rate larger than a given rate $\omega>0$, i.e. we want to solve the $\omega$-stabilization problem for the heat equation with memory. We first show that the spectrum of the state operator $A$ associated with this equation has an accumulation point at $-\omega_0<0$. Given a $\omega\in(0,\omega_0)$, we show that the $\omega$-stabilization problem for the heat equation with memory is solvable provided certain verifiable conditions on the control operator $B$ associated with this equation hold. We then consider an appropriate LQR problem for the heat equation with memory. For each $n\in\nline$, we construct finite-dimensional approximations $A_n$ and $B_n$ of $A$ and $B$, respectively, and then show that by solving a corresponding approximation of the LQR problem a feedback operator $K_n$ can be computed such that all the eigenvalues of $A_n + B_n K_n$ have real part less than $-\omega$. We prove that $K_n$ for $n$ sufficiently large solves the $\omega$-stabilization problem for the heat equation with memory. A crucial and nontrivial step in our proof is establishing the uniform (in $n$) stabilizability of the pair $(A_n+\omega I, B_n)$. We have validated our theoretical results numerically using two examples: an 1D example on a unit interval and a 2D example on a square domain.
}}
\end{center}

\noindent
{\bf Keywords}. Bilinear form, Galerkin approximation, Hautus test, linear quadratic regulator problem, Riccati equation, uniform stabilizability \vspace{-3mm}

%%%%%%%%%%**********%%%%%%%%%%**********%%%%%%%%%%**********%%%%%%%%%%
%%%%%%%%%%**********%%%%%%%%%%**********%%%%%%%%%%**********%%%%%%%%%%
\section{Introduction} \label{section: Introduction} \vspace{-1mm}
\setcounter{equation}{0} % Section 1

\ \ \ Let $\Omega \subset \rline^d$ with $d\in\nline$ be an open connected set with $C^2$-boundary $\partial \Omega$. Let $L^2(\Omega)$ be the usual Hilbert space of real-valued square integrable functions on $\Omega$. Define $\Qscr=\Omega \times (0,\infty)$ and $\Sscr =\partial \Omega \times (0,\infty)$. The heat equation with memory considered in this work is
\begin{align}
 &y_t(\xi,t)-\eta \Delta y(\xi,t)-\int_{0}^{t}e^{-\kappa(t-s)}\Delta y(\xi,s) \dd s = \sum_{i=1}^m b_i(\xi ) u_i(t) \ \ \quad\forall\, (\xi ,t)\in \Qscr, \label{eq:HM_PIDE} \\
 &y(\xi ,t)=0 \ \ \quad \forall\,(\xi,t)\in \Sscr. \label{eq:HM_BC}
\end{align}
Here $y(\cdot ,t)\in L^2(\Omega)$ is the state, $u_i(t)\in\rline$ and $b_i\in L^2(\Omega)$ for $i\in\{1,2,\ldots m\}$ are the control inputs and input shape functions, respectively, and $\eta$ and $\kappa$ are some positive constants. The integral term in \eqref{eq:HM_PIDE} depends explicitly on all the past values of $y$ and can be regarded as a memory term. In the absence of this term, \eqref{eq:HM_PIDE}-\eqref{eq:HM_BC} is the standard heat equation with interior control and Dirichlet boundary conditions. When the control inputs are zero, for every initial state $y(0)\in L^2(\Omega)$, the solution $y$ of \eqref{eq:HM_PIDE}-\eqref{eq:HM_BC} decays exponentially to zero with a certain (possibly small) decay rate. In this work, our objective is to numerically construct a state feedback controller for \eqref{eq:HM_PIDE}-\eqref{eq:HM_BC} such that the solution $y$ of the closed-loop system, for each initial state $y(0)\in L^2(\Omega)$, decays exponentially to zero with a larger decay rate $\omega$, i.e. we want to solve the $\omega$-stabilization problem for \eqref{eq:HM_PIDE}-\eqref{eq:HM_BC}. Under some conditions on the input shape functions we will show that this problem can be solved for any desired $\omega<\kappa+\frac{1}{\eta}$ and then construct a controller which solves the problem. The heat equation with memory in \eqref{eq:HM_PIDE}-\eqref{eq:HM_BC} is the linearization of a viscous Burgers equation with memory around its zero steady state, see \cite{AkMi:2022}, and the results developed in this paper are relevant to the local stabilization of the viscous Burgers equation with memory.

Several works in the literature have studied the controllability of heat equations with memory, see \cite{BaIa:2000}, \cite{GuIm:13}, \cite{HaPa:12}, \cite{IvPa:09}, \cite{SiZhZu:17}, \cite{ZhGa:14}. Approximate controllability of heat equations with memory, defined via completely monotone kernels, is established in \cite{BaIa:2000}. In \cite{IvPa:09} it is shown that one-dimensional heat equations with memory are not null controllable for large classes of memory kernels and controls. Even more, it is shown in \cite{GuIm:13} and \cite{HaPa:12} that certain heat equations with memory cannot even be steered to the zero state at desired time instants using boundary control. A similar result for heat equations with memory driven by interior controls is established in \cite{ZhGa:14}, which also shows that these equations are however approximately controllable. The above works which use interior control assume that the location of the control is fixed. Using moving interior control, a null controllability result for heat equations with memory is established in \cite{SiZhZu:17}.

Given the general lack of null controllability of heat equations with memory, more recent works have focussed on the stabilizability of these equations, see \cite{AkMi:2022}, \cite{BrKu:14}, \cite{BrKu:17}, \cite{LiZhGu:18}, \cite{Mun:15}. In \cite{BrKu:14}, a stabilizing feedback controller is obtained for a heat equation with memory by solving a Riccati equation corresponding only to the heat equation (without the memory term). For the same heat equation with memory, a reduced-order stabilizing compensator is constructed in \cite{BrKu:17} by solving Riccati equations for finite-dimensional approximations of the heat equation with memory. While \cite{BrKu:14}, \cite{BrKu:17} considered interior controls, stabilizing controllers for boundary controlled heat equations with memory are derived in \cite{LiZhGu:18}, \cite{Mun:15} without using Riccati equations. More recently in \cite{AkMi:2022}, a state feedback controller was designed to address the $\omega$-stabilization problem for a heat equation with memory (different from those considered in \cite{BrKu:14}, \cite{BrKu:17}, \cite{LiZhGu:18}, \cite{Mun:15}) by solving a finite-dimensional Riccati equation. In this work, we solve a similar $\omega$-stabilization problem by constructing a state feedback controller using the techniques from \cite{BaKu:1984} (see below for more details). The results in this work are extremely pertinent to the construction of reduced-order output feedback compensators for the $\omega$-stabilization problem of interest using ideas from \cite{Cur:03}, see Section \ref{sec6} for a brief discussion. We remark that the heat equations with memory in \cite{AkMi:2022}, \cite{BrKu:14}, \cite{BrKu:17}, \cite{Mun:15} are obtained by linearizing certain nonlinear equations and it is shown in these works %\cite{AkMi:2022}, \cite{BrKu:14}, \cite{BrKu:17}, \cite{Mun:15}
that stabilizing controllers developed for the former equations locally stabilize the latter equations.

The early lumping approach to the design of controllers for partial differential equation (PDE) models is as follows: (i) Approximate the PDE model with an ordinary differential equation (ODE) model obtained by approximating the spatial derivatives in the PDE, (ii) design finite-dimensional controllers for the ODE model and (iii) show that a controller designed for a sufficiently accurate ODE approximation of the PDE model solves the control problem for the PDE model satisfactorily. This approach has been used to design adaptive controllers for heat equations using the backstepping technique \cite{BaKr:2002}, \cite{BoBaKr:2003}, to construct control signals for motion planning of heat equations using the flatness technique \cite{ChNa:2025}, \cite{UtMeKu:2010}, and to design stabilizing controllers for PDEs by solving linear quadratic regulator (LQR) problems \cite{AkMiNaRa:2024}, \cite{BaRa:2024}, \cite{BaRa:2024b}, \cite{BuCh:2022}, \cite{GiRoTa:1992}, \cite{GrMo:1996}, \cite{KaMo:2013}, \cite{RaTaTu:2007}, \cite{SiChNa:2022}. There are two sets of results, one developed in \cite{BaKu:1984} (see also \cite{BaIt:1988}) and another in \cite[Chapter 4]{LaTr:2000a}, %(see also \cite{LaTr:1987}, \cite{LaTr:1986}),
which can be used to establish the applicability of the early lumping approach for solving LQR problems for a given PDE. While the results in \cite[Chapter 4]{LaTr:2000a} are stronger (they provide order of convergence associated with the early lumping approach), they also require more stringent hypothesis to hold which we are unable to verify for the heat equation with memory considered in this paper. On the other hand, the results in \cite{BaKu:1984} hold under weaker hypothesis which we are able to verify and so we use them in this paper.

In this article, we address the $\omega$-stabilization problem for the heat equation with memory \eqref{eq:HM_PIDE}-\eqref{eq:HM_BC} by solving a LQR problem via the early lumping approach. For this, we first rewrite \eqref{eq:HM_PIDE}-\eqref{eq:HM_BC} as an abstract evolution equation on the state space $L^2(\Omega)\times H^1_0(\Omega)$ with state operator $A$ and control operator $B$, see Section \ref{sec2}. In Theorem \ref{thm:Aspectrum} we show that the spectrum of $A$ consists of two sequences of eigenvalues, all them having a negative real part, with one sequence converging to $-\infty$ and the other to $-\omega_0=-\kappa-\frac{1}{\eta}$. We then associate a bilinear form $\mathfrak{a}$ with $A$ and show that $A$ generates an analytic semigroup on the state space. Using the Hautus test we derive conditions on $B$ in Theorem \ref{thm:ABstab} under which the pair $(A+\omega I, B)$ is stabilizable for a given $\omega\in (0,\omega_0)$.
%We next focus on the main objective of this paper, which is to compute a feedback controller which solves the $\omega$-stabilization problem for \eqref{eq:HM_PIDE}-\eqref{eq:HM_BC} by using finite-dimensional approximations $A_n$ and $B_n$ of $A$ and $B$, respectively.
We next focus on the main objective of this paper, which is to numerically compute a feedback controller which stabilizes $(A+\omega I, B)$ by solving an appropriate LQR problem using finite-dimensional approximations $A_n$ and $B_n$ of $A$ and $B$, respectively. (Note that $A_n$ and $B_n$ are defined for each $n\in\nline$ and the dimension of $A_n$ increases with $n$.) To address this objective we closely follow the ideas developed in \cite{BaKu:1984}. Central to applying the techniques in \cite{BaKu:1984} to our problem is establishing the uniform (in $n$) stabilizability of the pair $(A_n+\omega I, B_n)$. This uniform stabilizability does not follow directly via the arguments in \cite{BaKu:1984} and establishing it requires the more elaborate analysis presented in Propositions \ref{pr:approxeig} and \ref{pr:unifstab}. The reason for this is, unlike in \cite{BaKu:1984}, the state operator $A$ in this paper does not have a compact resolvent, see Remark \ref{rm:compactness} for more details. We remark that the heat equation with memory can be regarded as a coupled system consisting of a parabolic PDE and an ODE, see Section \ref{sec2}. Applying the techniques developed in \cite{BaKu:1984} for parabolic PDEs to this coupled system is the key contribution this paper. Theorem \ref{thm:main_result} contains our main result on constructing the desired feedback controller. Briefly, it states that the feedback operator $K_n$ computed by solving a finite-dimensional approximation of the LQR problem posed for \eqref{eq:HM_PIDE}-\eqref{eq:HM_BC}, wherein the approximate LQR problem is defined using $A_n$ and $B_n$, solves the $\omega$-stabilization problem provided $n$ is sufficiently large.

The rest of this article is organized as follows. We introduce the statement of the $\omega$-stabilization problem for the heat equation with memory \eqref{eq:HM_PIDE}-\eqref{eq:HM_BC} formally in Section \ref{sec2}. The properties of the state operator $A$ and control operator $B$ associated with \eqref{eq:HM_PIDE}-\eqref{eq:HM_BC} are presented in Section \ref{sec3}. Our solution to the $\omega$-stabilization problem, which is based on solving finite-dimensional Riccati equations, is in Section \ref{sec4}. We illustrate our theoretical results numerically using two examples in Section \ref{sec5}. Section \ref{sec6} contains some concluding remarks.

%%%%%%%%%%%%%%%%%%%%%%%%%%%%%%----------%%%%%%%%%%%%%%%%%%%%%%%%%%%%%
\noindent
{\bf Notations and definitions}. For each $\omega\in\rline$, we define $\cline_{\omega}^+=\{\alpha\in \cline\m\big|\m \Re\alpha>\omega\}$ and $\cline_{\omega}^-=\{\alpha\in \cline\m\big|\m \Re\alpha<\omega\}$. Let $\overline{\cline_{\omega}^+}$ and $\overline{\cline_{\omega}^-}$ denote the closures of $\cline_{\omega}^+$ and $\cline_{\omega}^-$, respectively. Let $X$ and $Y$ be Hilbert spaces. We denote the inner product and norm on $X$ by $\langle \cdot,\cdot \rangle_{X}$ and $\|\cdot\|_{X}$, respectively. Elements of the product space $X \times Y$ are written as $\sbm{x\\ y}$ with $x \in X$ and $y \in Y$ and $X \times Y$ is a Hilbert space with the inner product  $\left\langle\sbm{x_1\\y_1},\sbm{x_2\\y_2}\right\rangle_{X \times Y}=\left\langle x_1, x_2\right\rangle_X+\left\langle y_1, y_2\right\rangle_Y$. The space of bounded linear operators from $X$ to $Y$ is a Banach space denoted by $\mathcal{L}(X, Y)$ with norm written as $\|\cdot\|_{\mathcal{L}(X, Y)}$. When $X=Y$ we write $\mathcal{L}(X)$ instead of $\mathcal{L}(X,X)$. We use $I$ to denote the identity operator on the appropriate space (which will be clear from the context). \vspace{-2mm}%For any linear operator $A$ on $X$, we denote its domain by $D(A)$.

%%%%%%%%%%**********%%%%%%%%%%**********%%%%%%%%%%**********%%%%%%%%%%
%%%%%%%%%%**********%%%%%%%%%%**********%%%%%%%%%%**********%%%%%%%%%%
\section{Problem statement} \label{sec2} \vspace{-1mm}
\setcounter{equation}{0} % Section 2

\ \ \ Recall the heat equation with memory \eqref{eq:HM_PIDE}-\eqref{eq:HM_BC} in which the constants $\kappa, \eta>0$ and the input shape functions $b_i$ are known. In this section, we formally introduce the $\omega$-stabilization problem for \eqref{eq:HM_PIDE}-\eqref{eq:HM_BC} addressed in this paper.

By defining $z(\xi ,t)=\int_{0}^{t} e^{-\kappa(t-s)}y(\xi ,s)\dd s$, we can rewrite \eqref{eq:HM_PIDE}-\eqref{eq:HM_BC} as a coupled system consisting of a parabolic PDE and an ODE as follows: \vspace{-2mm}
\begin{align}
 &y_t(\xi ,t)=\Delta\left(\eta y(\xi ,t)+z(\xi ,t)\right)+\sum_{i=1}^m b_i(\xi) u_i(t) \FORALL (\xi,t)\in \Qscr,\label{eq:HM_PDE}\\
 &z_t(\xi ,t)=-\kappa z(\xi ,t) + y(\xi ,t)  \FORALL (\xi ,t)\in \Qscr, \label{eq:HM_ODE}\\
 &y( \xi ,t)=0, \qquad z(\xi ,t)=0 \FORALL (\xi ,t)\in \Sscr. \label{eq:HM_BCyz}
\end{align}
Let $H^k(\Omega)$ denote the usual real Sobolev space of order $k$. Let $H^1_0(\Omega)$ be the set of all functions in $H^1(\Omega)$ with trace zero. Then $H^1_0(\Omega)$ is a Hilbert space with the norm
$$  \|f\|_{H^1_0(\Omega)}=\Big(\sum_{j=1}^{d} \|D_{\xi_j}f\|^2_{L^2(\Omega)} \Big)^{\frac{1}{2}} \FORALL f\in H^1_0(\Omega). $$
Here $D_{\xi_j}f$ is the weak derivative of $f$ with respect to the $j^{\rm th}$-coordinate of $\rline^d$. Consider the Hilbert spaces $X=L^2(\Omega)\times H^1_0(\Omega)$ and $V=H^1_0(\Omega)\times H^1_0(\Omega)$. Clearly $V$ is a dense subset of $X$. The coupled system \eqref{eq:HM_PDE}-\eqref{eq:HM_BCyz} can be written as an abstract evolution equation on the state space $X$ as follows:
\begin{equation} \label{eq:HM_abs}
 \bbm{\dot{y}(t)\\ \dot{z}(t)}=A\bbm{y(t)\\z(t)}+Bu(t) \FORALL t>0.
\end{equation}
Here the state operator $A:D(A)\subset X\to X$ has domain %maps the domain $D(A)\subset X$ to $X$ with
\begin{equation} \label{eq:Adom}
 D(A)=\left\{ \bbm{y\\z}\in V \ \bigg{|} \ [\eta y+z]\in H^2(\Omega)\right\}
\end{equation}
with
\begin{equation} \label{eq:Adef}
 A\bbm{y \\ z}=\bbm{\Delta [\eta y+z]\\ y-\kappa z} \FORALL \bbm{y\\z}\in D(A),
\end{equation}
the input $u(t)=\bbm{u_1(t) & u_2(t) \ \cdots \ u_m(t)}^\top \in \rline^m$
and the control operator $B\in \Lscr(\rline^m,X)$ is defined as
\begin{equation} \label{eq:Bdef}
 B \alpha=\bbm{\sum_{i=1}^m b_i \alpha_i \\ 0} \FORALL \alpha\in \rline^m,
\end{equation}
where $\alpha_i$ is the $i^{\rm th}$-component of $\alpha$. The state operator $A$ generates an analytic semigroup $\tline$ on $X$, see Theorem \ref{thm:Aanalytic} in Section \ref{sec3.2}. For each input $u\in L^2((0,\infty);\rline^m)$ and initial state $\sbm{y^0 \\ z^0}\in X$, the mild solution
$\sbm{y \\ z}\in C([0,\infty);X)$ of \eqref{eq:HM_abs} is given by
\begin{equation} \label{eq:mild_solution}
 \bbm{y(t) \\ z(t)} =\tline(t) \bbm{y^0\\ z^0} + \int_0^t \tline(t-s) B u(s)\dd s \FORALL t\geq0.
\end{equation}
The above mild solution is the unique weak solution of \eqref{eq:HM_PDE}-\eqref{eq:HM_BCyz} corresponding to the input
$u$ and initial state $\sbm{y^0 \\ z^0}$, see for instance \cite[Proposition 4.2.5 and Remark 4.2.6]{obs_book}. We regard the first component $y$ of the above mild solution corresponding to some input $u$ and initial state $\sbm{y^0 \\ 0}$ to be the solution of \eqref{eq:HM_PIDE}-\eqref{eq:HM_BC} corresponding to the same input $u$ and initial state $y^0$.

Define $\omega_0=\kappa+\frac{1}{\eta}$. We address the following $\omega$-stabilization problem in this paper.
\begin{framed} \vspace{-2mm}
\begin{problem} \label{prob:wstab}
 Given $0<\omega<\omega_0$, numerically compute a stabilizing controller gain $K_\omega \in \Lscr(X,\rline^m)$ such that the strongly continuous semigroup $\tline^{cl}$ on $X$ generated by $A+ BK_\omega$ satisfies
\begin{equation} \label{eq:probbnd}
  \|\tline^{cl}(t)\|_{\Lscr(X)} \leq M e^{-(\omega+\epsilon) t} \FORALL t>0
\end{equation}
and some $M, \epsilon>0$.
\end{problem}
\end{framed}
\vspace{-2.5mm}

To solve the above problem, we choose an appropriate LQR problem for \eqref{eq:HM_abs}. For each $n\in\nline$, we construct finite-dimensional approximations $A_n$ and $B_n$ of $A$ and $B$, respectively, and then show that by solving a corresponding approximation of the chosen LQR problem a feedback operator $K_n$ can be computed such that the eigenvalues of $A_n + B_n K_n$ have real part less than $-\omega$. We prove that $K_\omega=K_n$ for $n$ sufficiently large solves the above problem, see Theorem \ref{thm:main_result} for more details. We remark that $K_n$ can be computed by solving a finite-dimensional Riccati equation numerically.

Note that $\tline^{cl}$ in Problem \ref{prob:wstab} is the semigroup associated with the closed-loop system obtained by applying the state-feedback control law $u=K_\omega\sbm{y \\ z}$ in \eqref{eq:HM_abs}. Suppose that \eqref{eq:probbnd} holds. Then the same control law when applied in \eqref{eq:HM_PIDE}-\eqref{eq:HM_BC} ensures that the trajectory $y$ of \eqref{eq:HM_PIDE}-\eqref{eq:HM_BC} satisfies $\|y(t)\|_{L^2(\Omega)}\leq M_0 e^{-(\omega+\epsilon)t} \|y(0)\|_{L^2(\Omega)}$ for some $M_0>0$. We remark that Problem \ref{prob:wstab} does not have a solution if $\omega\geq \omega_0$, see Remark \ref{rm:optimal}. \vspace{-2mm}

%%%%%%%%%%**********%%%%%%%%%%**********%%%%%%%%%%**********%%%%%%%%%%
%%%%%%%%%%**********%%%%%%%%%%**********%%%%%%%%%%**********%%%%%%%%%%
\section{Properties of the state operator $A$ and control operator $B$ \vspace{-2mm}} \label{sec3}
\setcounter{equation}{0} % Section 3

\ \ \ In Section \ref{sec3.1} we derive some spectral properties of the state operator $A$ introduced in \eqref{eq:Adom}-\eqref{eq:Adef}. In Section \ref{sec3.2} we associate a bilinear form with $A$ and show that $A$ generates an analytic semigroup on $X=L^2(\Omega)\times H^1_0(\Omega)$. We establish conditions for the stabilizability of the pair $(A+\omega I, B)$ for $0<\omega<\kappa+\frac{1}{\eta}$ using the Hautus test in Section \ref{sec3.3}. \vspace{-3mm}

%%%%%%%%%%**********%%%%%%%%%%**********%%%%%%%%%%**********%%%%%%%%%%
%%%%%%%%%%**********%%%%%%%%%%**********%%%%%%%%%%**********%%%%%%%%%%
\subsection{Spectral properties of $A$ \vspace{-1mm}} \label{sec3.1} \vspace{-1mm}% Section 3.1

\ \ \ Recall the following result about the eigenvalues and eigenfunctions of the Dirichlet Laplacian \cite[Theorem 3.6.1]{Ke:2015}:\vspace{-1mm} 

%%%%%%%%%%**********%%%%%%%%%%**********%%%%%%%%%%**********%%%%%%%%%%%
%------------------------- Begin theorem --------------------------%
\begin{framed} \vspace{-2mm}
\begin{theorem}\label{thm:Laplacian}
There exist a nondecreasing unbounded sequence of positive real eigenvalues $(\lambda_j)_{j\in\nline}$ and a sequence of eigenfunctions $(\psi_j)_{j\in\nline}$ in $H^1_0(\Omega)\cap C^{\infty}(\Omega)$ which form an orthonormal basis for $L^2(\Omega)$ such that \vspace{-1mm}
$$ -\Delta\psi_j = \lambda_j \psi_j \quad \textrm{in} \quad \Omega \vspace{-1mm} $$
for all $j\in\nline$. The dimension of the eigenspace of each $\lambda_j$ is finite. \vspace{-2mm}
%Furthermore, the sequence of functions $(\lambda_j^{-\frac{1}{2}} \psi_j)_{j\in\nline}$ form an orthonormal basis for $H^1_0(\Omega)$.
\end{theorem}
\end{framed}
%-------------------------- End Theorem ---------------------------%
%%%%%%%%%%**********%%%%%%%%%%**********%%%%%%%%%%**********%%%%%%%%%%%

\vspace{-1mm}
The next theorem about the spectral properties of $A$ is based on the results in \vspace{-1mm} \cite{AkMi:2022}.

%%%%%%%%%%**********%%%%%%%%%%**********%%%%%%%%%%**********%%%%%%%%%%%
%------------------------- Begin theorem --------------------------%
\begin{framed} \vspace{-3mm}
\begin{theorem}\label{thm:Aspectrum}
The state operator $A$ has two sequences of eigenvalues  $(\mu^+_j)_{j\in \nline}$ and $(\mu^-_j)_{j\in \nline}$ given by \vspace{-1mm}
\begin{equation} \label{eq:Aeigenvalues}
 \mu_j^{\pm}=\frac{-\left(\kappa+\eta \lambda_j\right) \pm \sqrt{\left(\kappa+\eta \lambda_j\right)^2-4 \lambda_j(1+\eta \kappa)}}{2},
\end{equation}
where $\lambda_j$ is the eigenvalue of $-\Delta$ introduced in Theorem \ref{thm:Laplacian}. Each of these eigenvalues has finite algebraic multiplicity.

Let $\omega_0=\kappa+\frac{1}{\eta}$. The real part of each of the eigenvalues in \eqref{eq:Aeigenvalues} is negative and
\begin{equation} \label{eq:Aeigaccu}
 \lim_{j\to \infty}\mu_j^+ =-\omega_0, \qquad  \lim_{j\to \infty}\mu_j^- =-\infty. \vspace{-1mm}
\end{equation}
Furthermore, for each $\epsilon>0$ the spectrum of $A$ contained in $\cline^+_{-\omega_0+\epsilon}$ consists only of the finitely many eigenvalues of $A$ which lie in $\cline^+_{-\omega_0+\epsilon}$. \vspace{-1mm}
\end{theorem}
\end{framed}
%-------------------------- End Theorem ---------------------------%
%%%%%%%%%%**********%%%%%%%%%%**********%%%%%%%%%%**********%%%%%%%%%%%

%%%%%%%%%%%%%%%%%%%%%%%%%%%%-- Proof --%%%%%%%%%%%%%%%%%%%%%%%%%%%%%
\begin{proof}
For finding the eigenvalues of $A$, we consider the equation \vspace{-1mm}
\begin{equation}\label{eq:eigA}
  A\bbm{y\\z}=\mu \bbm{y\\z} \vspace{-1mm}
\end{equation}
with $\mu\in\cline$ and $\sbm{y\\z}\in D(A)$. Using the definition of $A$ and introducing the notation $w=\eta y+z$, the above equation can be rewritten as \vspace{-1mm}
\begin{align}
  -\Delta w + \frac{\mu(\mu+\kappa)}{\eta(\mu+\kappa)+1}w =&\, 0 \quad\textrm{in}\quad \Omega,\qquad w\big|_{\partial\Omega}=0, \label{eq:eigAres1}\\
  \left(\eta(\mu+\kappa)+1\right)z-w=&\,0 \quad\textrm{in}\quad \Omega. \label{eq:eigAres2} \\[-4ex]\nonumber
\end{align}
From \eqref{eq:eigAres1} it is clear that $\mu$ is an eigenvalue of $A$ if and only if  $\frac{\mu(\mu+\kappa)}{\eta(\mu+\kappa)+1}$ is an eigenvalue of $\Delta$ (with Dirichlet boundary conditions). It now follows from Theorem \ref{thm:Laplacian} that we can find all the eigenvalues of $A$ by solving the equation \vspace{-1mm}
\begin{equation} \label{eq:eigrel}
 \frac{\mu(\mu+\kappa)}{\eta(\mu+\kappa)+1}=-\lambda_j \vspace{-1mm}
\end{equation}
for each $j\in\nline$. Solving the above quadratic equation in $\mu$ for each $j\in\nline$ we get the two sequences of eigenvalues  $(\mu^+_j)_{j\in \nline}$ and $(\mu^-_j)_{j\in \nline}$ given in \eqref{eq:Aeigenvalues}.

Recall that the algebraic multiplicity of an eigenvalue $\mu$ of $A$ is \vspace{-1mm}
$$\dim\Big(\bigcup_{k\in\nline}\,\ker (\mu I - A)^k\Big). \vspace{-3mm}$$
Let $\tilde X=L^2(\Omega)\times L^2(\Omega)$. Consider the operator $\tilde A:\tilde X\to \tilde X$ with domain \vspace{-1mm}
\begin{equation} \label{eq:tilAdom}
 D(\tilde A)=\left\{ \bbm{y\\z}\in \tilde X \ \bigg{|} \ [\eta y+z]\in H^2(\Omega)\cap H^1_0(\Omega)\right\}
\end{equation}
with
\begin{equation} \label{eq:tilAdef}
 \tilde A\bbm{y \\ z}=\bbm{\Delta [\eta y+z]\\ y-\kappa z} \FORALL \bbm{y\\z}\in D(\tilde A).
\end{equation}
Note that $\tilde A$ is the state operator $A$ in \eqref{eq:Adom}-\eqref{eq:Adef}, but on a larger space and with a larger domain. In particular, $D(A)\subset D(\tilde A)$ and $Av=\tilde A v$ for all $v\in D(A)$. Consequently, (i) the set of eigenvalues of $\tilde A$ contain the two sequences $(\mu^+_j)_{j\in \nline}$ and $(\mu^-_j)_{j\in \nline}$ given in \eqref{eq:Aeigenvalues} and (ii) if $v\in \ker(\mu I- A)^k$ for some $\mu\in \cline$ and $k\in\nline$, then $v\in \ker(\mu I- \tilde A)^k$. In \cite[Section 3.2]{AkMi:2022} it has been established that each of the eigenvalues of $\tilde A$ has finite algebraic multiplicity. It now follows from (i) and (ii) along with the definition of algebraic multiplicity that the eigenvalues $\mu_j^{\pm}$ of $A$ in \eqref{eq:Aeigenvalues} also have finite algebraic multiplicity.

From \eqref{eq:Aeigenvalues} it is obvious that the real part of the eigenvalues of $A$ is negative. The limits in \eqref{eq:Aeigaccu} can be established using series expansion \cite[Proposition 3.3(b)]{AkMi:2022}.

The spectrum of $\tilde A$ is the set
$$ \sigma(\tilde A) = \textrm{closure}\big\{(\mu^+_j)_{j\in \nline},\m (\mu^-_j)_{j\in \nline}\big\},$$
see \cite[Theorem 3.9]{AkMi:2022}. Fix $\mu\notin \sigma(\tilde A)$, i.e. $\mu$ is in the resolvent of $\tilde A$. Then for every $\sbm{f\\g}\in X$ there exists a unique $\sbm{y\\z}\in D(\tilde A)$ such that
$$ (\mu I -\tilde A)\bbm{y\\z} =\bbm{f\\g}, $$
which means that $(\eta y+z) \in H^2(\Omega) \cap H^1_0(\Omega)$ and $z\in L^2(\Omega)$ are unique functions which satisfy
\begin{align}
 -\Delta [\eta y+z] + \frac{\mu(\mu+\kappa)}{\eta(\mu+\kappa)+1}[\eta y+ z] =&\, f+ \frac{\mu }{\eta(\mu+\kappa)+1}g, \label{eq:res1pf}\\
 (\eta(\mu+\kappa)+1)z-[\eta y+z]=&\,\eta g.\label{eq:res2pf}
\end{align}
Applying \cite[Section 6.3, Theorem 4]{Ev:1988} (and the remark below it) to \eqref{eq:res1pf} we get
\begin{equation}\label{eq:estpf}
 \|\eta y+ z\|_{H^2(\Omega)} \leq \tilde C (\|f\|_{L^2(\Omega)} + \|g\|_{H^1_0(\Omega)})
\end{equation}
for some $\tilde C>0$ independent of $\sbm{f \\ g}$. Noting that $(\eta y+z), g\in H^1_0(\Omega)$, it follows from \eqref{eq:res2pf} that $y, z\in H^1_0(\Omega)$ and using \eqref{eq:res2pf} and \eqref{eq:estpf} we get
\begin{equation}\label{eq:estres}
 \|y\|_{H^1_0(\Omega)} + \|z\|_{H^1_0(\Omega)} \leq C (\|f\|_{L^2(\Omega)} + \|g\|_{H^1_0(\Omega)})
\end{equation}
for some $C>0$ independent of $\sbm{f \\ g}$. From the above discussion we can now conclude that for every $\sbm{f\\g}\in X$ there exists a unique $\sbm{y\\z}\in D(A)$ such that
$$ (\mu I -A)\bbm{y\\z} =\bbm{f\\g}$$
and
$$ \left\|\bbm{y \\ z}\right\|_X \leq \sqrt{2} C \left\|\bbm{f \\ g}\right\|_X, $$
which means that $\mu$ is also in the resolvent set of $A$. Hence the spectrum of $A$ is the set
$$ \sigma(A) = \textrm{closure}\big\{(\mu^+_j)_{j\in \nline},\m (\mu^-_j)_{j\in \nline}\big\},$$
which immediately implies the `Furthermore' part of the theorem.
\end{proof}
%%%%%%%%%%%%%%%%%%%%%%%%%-- End Proof --%%%%%%%%%%%%%%%%%%%%%%%%%%%%

%%%%%%%%%%%%%%%%%*******************************%%%%%%%%%%%%%%%%%%%%
\begin{remark} \label{rm:optimal}
Suppose that $\omega\geq\omega_0$. Theorem \ref{thm:Aspectrum} implies that $A+\omega I$ has infinitely many eigenvalues in $\overline{\cline^+_0}$. Since $B\in\Lscr(\rline^m, X)$ is a  finite-rank operator, it follows from \cite[Theorem 5.2.6]{CuZw:1995} that there does not exist a $K_\omega\in \Lscr(X,\rline^m)$ such that $A+\omega I + B K_\omega$ is exponentially stable, i.e. Problem \ref{prob:wstab} does not have a solution.
\end{remark}
%%%%%%%%%%%%%%%%%*******************************%%%%%%%%%%%%%%%%%%%%

%%%%%%%%%%**********%%%%%%%%%%**********%%%%%%%%%%**********%%%%%%%%%%
%%%%%%%%%%**********%%%%%%%%%%**********%%%%%%%%%%**********%%%%%%%%%%
\subsection{$A$ generates an analytic semigroup} \label{sec3.2} % Section 3.2

\ \ \ Recall that $V=H^1_0(\Omega)\times H^1_0(\Omega)$. Consider the bilinear form $\mathfrak{a}:V\times V \to \rline$ defined as
\begin{equation} \label{eq:sesqA}
 \mathfrak{a}\left(\bbm{y\\z},\bbm{p\\q}\right)= \big\langle [\eta y+z], p \big\rangle_{H^1_0(\Omega)} - \big\langle [y-\kappa z],q \big\rangle_{H^1_0(\Omega)} \FORALL \bbm{y\\z},\bbm{p\\q}\in V.
\end{equation}
A simple calculation gives
\begin{equation}\label{eq:sesqAcont}
 \big|\mathfrak{a}(v_1, v_2)\big| \leq (2+\eta+\kappa) \|v_1\|_V \|v_2\|_V \FORALL v_1, v_2\in V,
\end{equation}
\begin{equation}\label{eq:sesqAcoer}
 \mathfrak{a}(v,v) \geq \min \{\eta,\kappa\}\|v\|^2_V \FORALL v\in V.
\end{equation}
The operator $\hat A$ associated with the bilinear form $\mathfrak{a}$ is defined as follows:
$$ D(\hat A) = \{u\in V \m\big|\m \mathfrak{a}(u,v)=\langle w,v \rangle_X \textrm{ for some } w\in X \textrm{ and all } v\in V\} $$
and for $u\in D(\hat A)$ and $w$ such that $\mathfrak{a}(u,v)=\langle w,v \rangle_X$ for all $v\in V$ we let $\hat A u = -w$. In the next theorem we show that $\hat A$ is the same as $A$ and it generates an analytic semigroup on $X$.

%%%%%%%%%%**********%%%%%%%%%%**********%%%%%%%%%%**********%%%%%%%%%%%
%------------------------- Begin theorem --------------------------%
\begin{framed} \vspace{-2mm}
\begin{theorem}\label{thm:Aanalytic}
The state operator $A$ in \eqref{eq:Adom}-\eqref{eq:Adef} is the operator associated with the bilinear form $\mathfrak{a}$ in \eqref{eq:sesqA} and it generates an analytic semigroup $\tline$ on $X$.  \end{theorem}
\end{framed}
%-------------------------- End Theorem ---------------------------%
%%%%%%%%%%**********%%%%%%%%%%**********%%%%%%%%%%**********%%%%%%%%%%%

%%%%%%%%%%%%%%%%%%%%%%%%%%%%-- Proof --%%%%%%%%%%%%%%%%%%%%%%%%%%%%%
\begin{proof}
The bilinear form $\mathfrak{a}$ is continuous since \eqref{eq:sesqAcont} holds, coercive since \eqref{eq:sesqAcoer} holds and $V$ is dense in $X$. Therefore from \cite[Part II, Chapter 1, Theorem 2.12]{BePrDeMi:2007} we get that the operator $\hat A$ associated with $\mathfrak{a}$ generates an analytic semigroup $\tline$ on $X$. 

Let $\sbm{y \\ z}\in D(A)$. Then for all $\sbm{p \\ q}\in V$ we have
\begin{align*}
 \left\langle -A \bbm{y \\ z}, \bbm{p \\ q} \right\rangle_X
 &= -\langle \Delta (\eta y + z), p \rangle_{L^2(\Omega)} - \langle y-\kappa z, q \rangle_{H^1_0(\Omega)} \\
 &= \sum_{j=1}^n \langle D_{\xi_j}(\eta y + z), D_{\xi_j} p \rangle_{L^2(\Omega)} - \langle y-\kappa z, q \rangle_{H^1_0(\Omega)}\\
 &=\mathfrak{a}\left(\bbm{y\\z},\bbm{p\\q}\right).
\end{align*}
It now follows from the definition of $\hat A$ that $\sbm{y \\ z}\in D(\hat A)$ and $\hat A \sbm{y \\ z}=A \sbm{y \\ z}$. Hence $D(A) \subset D(\hat A)$ and $Au=\hat A u$ for all $u\in D(A)$. Fix $\lambda>0$  belonging to the resolvent set of $\hat A$. From Theorem \ref{thm:Aspectrum} we get that $\lambda$ is also in the resolvent set of $A$. So clearly
$$ (\lambda I-\hat A)D(\hat A) = X = (\lambda I-A) D(A) = (\lambda I-\hat A) D(A).$$
Applying $(\lambda I-\hat A)^{-1}\in\Lscr(X)$ to the first and last terms above we get $D(A)=D(\hat A)$. We have already shown that $Au=\hat A u$ for all $u\in D(A)$. Therefore $A$ is same as the operator $\hat A$ associated with the bilinear form $\mathfrak{a}$ and it is the generator of the analytic semigroup $\tline$ on $X$.
\end{proof}
%%%%%%%%%%%%%%%%%%%%%%%%%%-- End Proof --%%%%%%%%%%%%%%%%%%%%%%%%%%%

%%%%%%%%%%**********%%%%%%%%%%**********%%%%%%%%%%**********%%%%%%%%%%
%%%%%%%%%%**********%%%%%%%%%%**********%%%%%%%%%%**********%%%%%%%%%%
\subsection{$\omega$-stabilizability of $(A, B)$ for $0<\omega<\omega_0$} \label{sec3.3} % Section 3.3

\ \ \ Recall the state operator $A$ and the constants $\kappa, \eta>0$ from \eqref{eq:Adom}-\eqref{eq:Adef}, the control operator $B$ from \eqref{eq:Bdef} and $\omega_0=\kappa+\frac{1}{\eta}$. The pair $(A, B)$ is said to be $\omega$-stabilizable if there exists a $K\in\Lscr(X,\rline^m)$ such that the semigroup $\tline^{cl}$ generated by $A+ BK$ satisfies \eqref{eq:probbnd} for some $M,\epsilon>0$. In Theorem \ref{thm:ABstab} we show that the pair $(A, B)$ is $\omega$-stabilizable for $0<\omega<\omega_0$ under a certain non-orthogonality assumption on $B$.

A simple calculation yields the following definition for the adjoint operator $A^*:D(A^*)\subset X \to X$: \vspace{-1mm}
\begin{equation} \label{eq:Adom-star}
 D(A^*)=\left\{ \bbm{y\\z}\in V \ \bigg{|} \ [\eta y-z]\in H^2(\Omega)\right\} \vspace{-1mm}
\end{equation}
with
\begin{equation} \label{eq:Adef-star}
 A^*\bbm{y \\ z}=\bbm{\Delta [\eta y-z]\\ -y-\kappa z} \FORALL \bbm{y\\z}\in D(A^*).
\end{equation}
It is easy to see that the adjoint operator $B^*\in \Lscr(X,\rline^m)$ is given by
\begin{equation} \label{eq:Bdef-star}
 B^*\bbm{y \\ z}=\bbm{\langle b_1, y\rangle_{L^2(\Omega)} & \langle b_2, y\rangle_{L^2(\Omega)} \ \ \cdots \ \ \langle b_m, y\rangle_{L^2(\Omega)}}^\top \FORALL \bbm{y\\z}\in X. \vspace{-1mm}
\end{equation}

Recall the eigenvalues $\lambda_j$ and eigenvectors $\psi_j$ of the Dirichlet Laplacian ($-\Delta$) from Theorem \ref{thm:Laplacian} and the eigenvalues $\mu_j^\pm$ of $A$ (which are expressed in terms of $\lambda_j$) from \eqref{eq:Aeigenvalues}. We now present conditions for the $\omega$-stabilizability of the pair $(A,B)$.

%%%%%%%%%%**********%%%%%%%%%%**********%%%%%%%%%%**********%%%%%%%%%%%
%------------------------- Begin theorem --------------------------%
\begin{framed} \vspace{-2mm}
\begin{theorem}\label{thm:ABstab}
Let $\omega_0=\kappa+\frac{1}{\eta}$. Fix $0<\omega<\omega_0$. Then the spectrum of $A$ lying in $\overline{\cline_{-\omega}^+}$ is the set $\{\mu_j^+ \m\big|\m \Re \mu_j^+ \geq -\omega \}\cup \{\mu_j^- \m\big|\m \Re \mu_j^- \geq -\omega \}$ which consists of finitely many eigenvalues  %$\{\mu_j^\pm \m\big|\m j=1,2,\ldots J\}\setminus \{\mu_J^- \m\big|\m \mu_J^-<-\omega\}$.
of finite algebraic multiplicity. Consider the finite set
$$ \Jscr = \{j\in \nline \m\big|\m \mu_j^+ \textrm{ and/or }\mu_j^- \textrm{ belong to } \overline{\cline_{-\omega}^+} \}.$$
Suppose that for each $j\in\Jscr$ and every $\phi\in H^2(\Omega)\cap H_0^1(\Omega)$ which satisfies $-\Delta \phi = \lambda_j\phi$ we have
\begin{equation}\label{eq:Borth}
  B^*\bbm{\phi \\ 0} \neq0.
\end{equation}
Then the pair $(A, B)$ is $\omega$-stabilizable. \vspace{-1mm}
\end{theorem}
\end{framed}
%-------------------------- End Theorem ---------------------------%
%%%%%%%%%%**********%%%%%%%%%%**********%%%%%%%%%%**********%%%%%%%%%%%

%%%%%%%%%%**********%%%%%%%%%%**********%%%%%%%%%%**********%%%%%%%%%%%
\begin{remark} \label{rm:checking}
From Theorem \ref{thm:Laplacian} we have that the eigenspace corresponding to the eigenvalue $\lambda_j$ of the Dirichlet Laplacian ($-\Delta$) is finite-dimensional. Suppose that the functions $\{\phi_1,\phi_2,\ldots \phi_k\}$ form a basis for the eigenspace. Then verifying \eqref{eq:Borth} for each  $\phi\in H^2(\Omega)\cap H_0^1(\Omega)$ which satisfies $-\Delta \phi = \lambda_j\phi$ is equivalent to verifying that the rank of the matrix \vspace{-1mm}
$$\bbm{B^*\bbm{\phi_1 \\ 0} & B^*\bbm{\phi_2 \\ 0}\ \ \cdots\ \  B^*\bbm{\phi_k \\ 0}} \vspace{-1mm}$$
is $k$. Hence to apply the above theorem and conclude that the pair $(A, B)$ is $\omega$-stabilizable it is sufficient to verify a matrix rank condition for each $j\in\Jscr$.
\end{remark}
%%%%%%%%%%**********%%%%%%%%%%**********%%%%%%%%%%**********%%%%%%%%%%%

%%%%%%%%%%%%%%%%%%%%%%%%%%%%-- Proof --%%%%%%%%%%%%%%%%%%%%%%%%%%%%
\begin{proof}
Let $\omega_0$ and $\omega$ be as in the theorem statement. Define \vspace{-1mm}
\begin{equation}\label{eq:Awdefn}
  A_\omega=A+\omega I. \vspace{-1mm}
\end{equation}
Since $A$ generates an analytic semigroup $\tline$ on $X$, see Theorem \ref{thm:Aanalytic}, it follows that $A_\omega$ generates an analytic semigroup $\tline_\omega$ on $X$. From Theorem \ref{thm:Aspectrum} we get that the spectrum of $A$ lying in $\overline{\cline_{-\omega}^+}$ is the set $\{\mu_j^+ \m\big|\m \Re \mu_j^+ \geq -\omega \}\cup \{\mu_j^- \m\big|\m \Re \mu_j^- \geq -\omega \}$ which consists of finitely many eigenvalues of finite algebraic multiplicity
and the rest of the spectrum of $A$ has real part less than $-\omega-\delta$ for some $\delta>0$. So clearly the spectrum of $A_\omega$ lying in $\overline{\cline_0^+}$ is the set
\begin{equation} \label{eq:sigwp}
 \sigma^+_\omega=\{\mu_j^+ + \omega \m\big|\m \Re \mu_j^+ \geq -\omega \}\cup \{\mu_j^-+\omega \m\big|\m \Re \mu_j^- \geq -\omega \} \vspace{-1mm}
\end{equation}
which consists of finitely many eigenvalues of finite algebraic multiplicity and the rest of the spectrum of $A_\omega$, denoted by $\sigma^-_\omega$, has real part less than $-\delta$. In particular, \vspace{-1mm}
\begin{equation} \label{eq:specdet}
 \lambda\in \sigma^-_\omega \implies \Re\lambda<-\delta. \vspace{-1mm}
\end{equation}
Let $\Pi^-_\omega$ be the spectral projection on $\sigma^-_\omega$. Define $X^-_\omega=\Pi^-_\omega X$ and let $A^-_\omega$ and $\tline^-_\omega$ be the restrictions of $ A_\omega$ and $\tline_\omega$, respectively, to $X^-_\omega$. From \cite[Lemma 2.5.7]{CuZw:1995} we get that the spectrum of $A^-_\omega$ is $\sigma^-_\omega$ and that $A^-_\omega$ is the generator of the strongly continuous semigroup $\tline^-_\omega$. Furthermore, $\tline^-_\omega$ is an analytic semigroup since it is the restriction of an analytic semigroup. Hence it satisfies the spectrum determined growth condition \cite[Part II, Chapter 1, Corollary 2.5]{BePrDeMi:2007} and so using \eqref{eq:specdet} we get \vspace{-1mm}
$$\|\tline^-_\omega(t)\|\leq C e^{-\delta t} \FORALL t\geq0 \vspace{-1mm}$$
and some $C>0$. From the above discussion it follows that the pair $(A_\omega, B)$ satisfies all the assumptions required for applying \cite[Part V, Chapter 1, Proposition 3.3]{BePrDeMi:2007}. We will now show that the second statement in that result holds, i.e. \vspace{-1mm}
\begin{equation}\label{eq:2ndstat}
  \ker(\lambda I - A_\omega^*) \cap \ker(B^*) = \{0\} \FORALL \lambda\in\sigma^+_\omega. \vspace{-1mm}
\end{equation}

Fix $\lambda\in\sigma^+_\omega$. Note that $A_\omega^*=A^*+\omega I$ and $D(A_\omega^*)=D(A^*)$. Suppose that \vspace{-1mm}
\begin{equation}\label{eq:KerAw}
  (\lambda I - A_\omega^*)\bbm{p\\q}=0 \vspace{-1mm}
\end{equation}
for some $\sbm{p\\q}\in D(A_\omega^*)$. Since $\lambda\in \sigma^+_\omega$ it follows from \eqref{eq:sigwp} that $\lambda=\mu+\omega$, where $\mu$ is an eigenvalue of $A$ with
$\mu=\mu_j^+$ or $\mu=\mu_j^-$ for some $j\in\Jscr$.
Substituting $\lambda=\mu +\omega$ and $A_\omega=A+\omega I$ in \eqref{eq:KerAw} and simplifying it using the definition of $A^*$ we get that $p,q\in H^2(\Omega)\cap H^1_0(\Omega)$, $p=-(\mu +\kappa)q$ and \vspace{-1mm}
$$ -\frac{\mu (\mu + \kappa)}{\eta(\mu +\kappa)+1}q+\Delta q=0. \vspace{-1mm}$$
Using \eqref{eq:eigrel} (and the definitions of $\mu_j^+$ and $\mu_j^-$) it follows that the coefficient in the above equation is $\lambda_j$ for some $j\in\Jscr$ and so we have $-\Delta q=\lambda_j q$. Since $\lambda_j>0$ (see Theorem \ref{thm:Laplacian}), looking at the coefficient again we get that $\mu+\kappa\neq0$. From this, the assumption in \eqref{eq:Borth} and the definition of $B^*$ we get \vspace{-1mm}
$$ B^*\bbm{p\\q} = B^*\bbm{p\\0} = -(\mu +\kappa)B^*\bbm{q\\0} \neq0. \vspace{-1mm}$$
So if $\lambda\in\sigma^+_\omega$ and $\sbm{p\\q}\in \ker(\lambda I-A_\omega^*)$, then $\sbm{p\\q}\notin \ker (B^*)$, i.e. \eqref{eq:2ndstat} holds.

We have shown that the pair $(A_\omega, B)$ satisfies all the assumptions required for applying \cite[Part V, Chapter 1, Proposition 3.3]{BePrDeMi:2007} and that the second statement in that result, i.e. \eqref{eq:2ndstat}, holds. Therefore the first statement in that result, which is equivalent to the second statement, also holds and the pair $(A_\omega, B)$ is stabilizable. So there exist $K\in\Lscr(X,\rline^m)$ such that the semigroup generated by $A_\omega+BK=A+\omega I + BK$ is exponentially stable, which clearly implies that the semigroup $\tline^{cl}$ generated by $A+BK$ satisfies \eqref{eq:probbnd} for some $M,\epsilon>0$. This completes the proof of the theorem. \vspace{-2mm}
\end{proof}
%%%%%%%%%%%%%%%%%%%%%%%%%%-- End Proof --%%%%%%%%%%%%%%%%%%%%%%%%%%%

%%%%%%%%%%**********%%%%%%%%%%**********%%%%%%%%%%**********%%%%%%%%%%
%%%%%%%%%%**********%%%%%%%%%%**********%%%%%%%%%%**********%%%%%%%%%%
\section{LQR controller design \vspace{-1mm}} \label{sec4} \setcounter{equation}{0} % Section 4

\ \ \ Recall the operators $A$ and $B$ from Section \ref{sec2}. Consider the abstract evolution equation
\begin{equation} \label{eq:HM_absomega}
 \bbm{\dot{y}(t)\\ \dot{z}(t)}=A_\omega\bbm{y(t)\\z(t)}+Bu(t) \FORALL t>0
\end{equation}
on the state space $X=L^2(\Omega)\times H^1_0(\Omega)$. Here $A_\omega= A+\omega I$ for some constant $\omega>0$. Note that \eqref{eq:HM_absomega} is obtained from \eqref{eq:HM_abs} by replacing $A$ with $A_\omega$. Since $A$ generates an analytic semigroup $\tline$ on $X$, it follows that $A_\omega$ is the generator of the analytic semigroup $\tline_\omega$ on $X$, where $\tline_\omega(t) = e^{\omega t}\tline(t)$ for all $t\geq0$. For every input $u\in L^2((0,\infty);\rline^m)$ and initial state $\sbm{y^0 \\ z^0}\in X$, the corresponding mild solution
$\sbm{y \\ z}\in C([0,\infty);X)$ of \eqref{eq:HM_absomega} is given by \vspace{-2mm}
\begin{equation} \label{eq:mildomega}
 \bbm{y(t) \\ z(t)} =\tline_\omega(t) \bbm{y^0\\ z^0} + \int_0^t \tline_\omega(t-s) B u(s)\dd s \FORALL t\geq0. \vspace{-1mm}
\end{equation}

Consider the following quadratic cost functional $J$ which is associated with \eqref{eq:HM_absomega} and maps $L^2((0,\infty);\rline^m) \times X$ to (possibly infinite) scalars:
\begin{equation}\label{eq:Cost}
  J\left(u,\bbm{y^0\\z^0}\right)=\int_{0}^{\infty} \left( \left\langle \bbm{y(t)\\z(t)}, Q\bbm{y(t)\\z(t)} \right\rangle_{X}+u^{\top}(t)Ru(t) \right) \dd t.
\end{equation}
Here $Q\in \Lscr(X)$ is a self-adjoint coercive operator and $R\in \rline^{m\times m}$ is a symmetric positive-definite matrix and $\sbm{y \\ z}$ is the mild solution of \eqref{eq:HM_absomega} corresponding to the input $u\in L^2((0,\infty);\rline^m)$ and initial state $\sbm{y^0 \\  z^0}\in X$. Clearly $Q$ and $R$ are boundedly invertible and have coercive square roots which are also boundedly invertible. The algebraic Riccati equation associated with the minimization of $J$ over all possible inputs is
\begin{equation} \label{eq:Riccati}
 A_\omega^{*} \Pi+\Pi A_\omega - \Pi B R^{-1} B^{*}\Pi + Q = 0.
\end{equation}
Recall that $\omega_0=\kappa+\frac{1}{\eta}$. Suppose that $0<\omega<\omega_0$ and $B$ satisfies the hypothesis in Theorem \ref{thm:ABstab}. It then follows from Theorem \ref{thm:ABstab} that there exists $K\in\Lscr(X,\rline^m)$ such that the semigroup generated by $A_\omega+BK$ is exponentially stable. The semigroup generated by $A_\omega+LQ^{\frac{1}{2}}$, where $Q^{\frac{1}{2}}$ is the coercive square root of $Q$ and $L=-\omega Q^{-\frac{1}{2}} \in\Lscr(X)$, is also exponentially stable since $A$ is exponentially stable. Therefore there exists a unique nonnegative solution $\Pi\in\Lscr(X)$ to \eqref{eq:Riccati} and $A_\omega-BR^{-1} B^*\Pi$ generates an exponentially stable semigroup, see \cite[Theorem 6.2.7]{CuZw:1995}, which means that $K_\infty=-R^{-1}B^*\Pi$ stabilizes the pair $(A+\omega I,B)$. In other words, a $K_\omega$ which solves Problem \ref{prob:wstab} can be obtained by computing the nonnegative solution $\Pi$ of \eqref{eq:Riccati}.

In this section, we present an approach for computing a good approximation of the nonnegative solution $\Pi$ of \eqref{eq:Riccati}. More specifically, in Section \ref{sec4.1} we derive a sequence of finite-dimensional approximations of the abstract linear system in \eqref{eq:HM_absomega} and of the cost function $J$ in \eqref{eq:Cost}, wherein the sequences are indexed by $n$, and then introduce the sequence (again indexed by $n$) of finite-dimensional algebraic Riccati equations associated with them. In Section \ref{sec4.2}, we prove that the sequence of finite-dimensional approximations of \eqref{eq:HM_absomega} are uniformly (in $n$) stabilizable and then conclude via a result from \cite{BaKu:1984} that the solutions $\{\Pi_n \big| n\in\nline\}$ of the sequence of finite-dimensional algebraic Riccati equations converge (strongly) to the nonnegative solution $\Pi$ of \eqref{eq:Riccati} as $n\to\infty$, see Theorem \ref{thm:main_result}. So $K_\omega=-R^{-1}B^*\Pi_n$ solves Problem \ref{prob:wstab} provided we take $n$ sufficiently large. Note that $\Pi_n$ can be computed easily and so the desired $K_\omega$ can also be computed easily, see the examples in Section \ref{sec5}.  \vspace{-1mm}

%%%%%%%%%%**********%%%%%%%%%%**********%%%%%%%%%%**********%%%%%%%%%%
%%%%%%%%%%**********%%%%%%%%%%**********%%%%%%%%%%**********%%%%%%%%%%
\subsection{Finite-dimensional approximations \vspace{0mm}} \label{sec4.1} % Section 4.1

\ \ \ Let $(V_n)_{n\in\nline}$ be a sequence of finite-dimensional subspaces of $V=H^1_0(\Omega)\times H^1_0(\Omega)$ with $V_n=H_n \times H_n$ for each $n\in\nline$, where $H_n$ is a finite-dimensional subspace of $H^1_0(\Omega)$. For each $n\in\nline$, we regard $V_n$ as a subspace of $X$ and endow it with the inner product and norm from $X$. Recall the bilinear form $\mathfrak{a}:V\times V\to\rline$ from \eqref{eq:sesqA} which is associated with the state operator $A$ in \eqref{eq:Adom}-\eqref{eq:Adef}. We obtain the finite-dimensional approximation $A_n$ of $A$ by restricting $\mathfrak{a}$ to $V_n\times V_n$, i.e. we determine $A_n\in\Lscr(V_n)$ via the following expression:
\begin{equation}\label{eq:A-approx}
 \langle -A_n v_1,v_2 \rangle_X = \mathfrak{a}(v_1,v_2) \FORALL v_1,v_2\in V_n.
\end{equation}
Let $P_n$ be the orthogonal projection operator from $X$ to $V_n$. 
Recall $B$ from \eqref{eq:Bdef}. The approximation $B_n\in\Lscr(\rline^m,V_n)$ of $B$ is defined as follows:
\begin{equation} \label{eq:B-aaprox}
 B_n \alpha=\sum_{i=1}^m \alpha_i P_n\bbm{ b_i \\ 0} \FORALL \alpha\in \rline^m,
\end{equation}
where $\alpha_i$ is the $i^{\rm th}$-component of $\alpha$. For each $n\in\nline$, using $A_n$ and $B_n$ we define the $n^{\rm th}$ finite-dimensional approximation of the abstract evolution equation \eqref{eq:HM_absomega} to be the linear ordinary differential equation
\begin{equation} \label{eq:approx system}
 \bbm{\dot{y_n}(t) \\ \dot{z_n}(t)} = A_{\omega,n} \bbm{y_n(t) \\ z_n(t)} + B_n u(t) \FORALL t>0
\end{equation}
on the state space $V_n$. Here $A_{\omega,n}= A_n+\omega I$. Clearly $y_n(t)$ and $z_n(t)$ belong to $H_n$.

Recall $Q$ and $R$ from \eqref{eq:Cost} and let $Q_n=P_n Q P_n$. Clearly $Q_n\in\Lscr(V_n)$ is a self-adjoint coercive operator. Consider the following quadratic cost functional $J_n$, which maps $L^2((0,\infty);\rline^m) \times V_n$ to (possibly infinite) scalars, associated with \eqref{eq:approx system}:
\begin{equation}\label{eq:Cost_approx}
  J_n\left(u,\bbm{y^0_n\\z^0_n}\right)=\int_{0}^{\infty} \left( \left\langle \bbm{y_n(t)\\z_n(t)}, Q_n\bbm{y_n(t)\\z_n(t)} \right\rangle_{X}+u^{\top}(t)Ru(t) \right) \dd t.
\end{equation}
Here $\sbm{y_n \\ z_n}$ is the solution of \eqref{eq:approx system} corresponding to input $u\in L^2((0,\infty);\rline^m)$ and initial state $\sbm{y_n^0 \\  z_n^0}\in V_n$. Note that $J_n$ can be regarded as a restriction of $J$ to the dynamics of \eqref{eq:approx system}. The algebraic Riccati equation associated with the minimization of $J_n$ over all possible inputs is
\begin{equation} \label{eq:Riccati_approx}
 A_{\omega,n}^{*} \Pi_n + \Pi_n A_{\omega,n} - \Pi_n B_n R^{-1} B_n^{*}\Pi_n + Q_n = 0.
\end{equation}
In Section \ref{sec4.2} we show that, under certain conditions, there exists a unique nonnegative solution $\Pi_n\in\Lscr(V_n)$ of \eqref{eq:Riccati_approx} and it converges strongly to the unique nonnegative solution $\Pi\in \Lscr(X)$ of \eqref{eq:Riccati_approx} as $n\to\infty$.

%%%%%%%%%%**********%%%%%%%%%%**********%%%%%%%%%%**********%%%%%%%%%%
%%%%%%%%%%**********%%%%%%%%%%**********%%%%%%%%%%**********%%%%%%%%%%
\subsection{Convergence of $\Pi_n$ to $\Pi$} \label{sec4.2} % Section 4.2

\ \ \ Let $0<\omega<\omega_0$. Suppose that $B$ satisfies the hypothesis in Theorem \ref{thm:ABstab}. Then there exists an operator $K\in\Lscr(X,\rline^m)$ such that $A_\omega+BK$ generates an exponentially stable semigroup on $X$, see Theorem \ref{thm:ABstab}. Fix such a $K$. Define
\begin{equation} \label{eq:lenovo}
A_\omega^K = A_\omega+BK, \qquad A_{\omega,n}^K = A_{\omega,n}+B_n K P_n.
\end{equation}
In Proposition \ref{pr:approxeig} we show that if $\lambda_n$ is an eigenvalue of $A_{\omega,n}^K$ for each $n$, then every accumulation point of the sequence $(\lambda_n)_{n\in\nline}$ is an eigenvalue of $A_{\omega}^K$. Using this result we establish in Proposition \ref{pr:unifstab} that the pair $(A_{\omega,n},B_n)$ is uniformly (in $n$) stabilizable. Finally, appealing to \cite[Theorem 2.2]{BaKu:1984} we conclude that the nonnegative solution $\Pi_n$ of \eqref{eq:Riccati_approx} converges to the nonnegative solution $\Pi$ of \eqref{eq:Riccati_approx} as $n\to\infty$. We also show that $K_\omega=-R^{-1}B_n^*\Pi_n P_n$ solves Problem \ref{prob:wstab} if $n$ is sufficiently large.

We require the approximating subspaces $V_n$ to satisfy the following natural assumption: \vspace{-2mm}
\begin{framed} \vspace{-2mm}
\begin{assumption}\label{as:subspace}
 For every $v\in V$, there exists a sequence $(v_n)_{n\in\nline}$ in $V$ with $v_n\in V_n$ for all $n\in\nline$ such that
 $$\lim_{n\to \infty}\|v_n-v\|_V=0. \vspace{-3mm}$$
\end{assumption}
\end{framed}
\vspace{-3mm}

The above assumption implies that \vspace{1mm}
\begin{equation}\label{eq:assumconv}
 \lim_{n\to \infty}\|P_n x - x\|_X=0 \FORALL x\in X. \vspace{10mm}
\end{equation}

Consider the bilinear form $\mathfrak{a}^{*}:V\times V\to \rline$ defined as
\begin{equation}\label{eq:sesq-Astar}
 \mathfrak{a}^{*} \left(\bbm{y\\z},\bbm{p\\q}\right) = \langle [\eta y -z],p \rangle_{H^1_0(\Omega)} + \langle [y+\kappa z],q \rangle_{H^1_0(\Omega)} \FORALL \bbm{y\\z},\bbm{p\\q}\in V.
\end{equation}
A simple calculation gives
\begin{equation} \label{eq:sesqAstarcont}
 \big|\mathfrak{a}^{*}(v_1, v_2)\big| \leq (2+\eta+\kappa) \|v_1\|_V \|v_2\|_V \FORALL v_1, v_2\in V,
\end{equation}
\begin{equation} \label{eq:sesqAstarcoer}
 \mathfrak{a}^{*}(v,v) \geq \min \{\eta,\kappa\}\|v\|^2_V \FORALL v\in V.
\end{equation}
Mimicking the proof of Theorem \ref{thm:Aanalytic} we can show that the operator associated with the bilinear form $\mathfrak{a}^{*}$ is the adjoint operator $A^{*}$ defined in
\eqref{eq:Adom-star}-\eqref{eq:Adef-star}. 
Next, define the bilinear form ${\mathfrak{a}_\omega^K}^*:V\times V\to \rline$ as follows:
\begin{equation}\label{eq:sesq-AstarK}
 {\mathfrak{a}_\omega^K}^*(v_1,v_2) = \mathfrak{a}^{*}(v_1,v_2)-\langle (BK)^{*}v_1,v_2 \rangle_X - \omega\langle v_1,v_2 \rangle_{X} \FORALL v_1,v_2\in V.
\end{equation}
Since $A^*$ is associated with $\mathfrak{a}^*$ and $BK\in\Lscr(X)$ it follows that the operator associated with the bilinear form ${\mathfrak{a}_\omega^K}^*$ is the adjoint operator $(A+BK+\omega I)^{*}$. Using the estimates \eqref{eq:sesqAstarcont} and \eqref{eq:sesqAstarcoer} we get
\begin{equation}\label{eq:a_Kstar1}
\big|{\mathfrak{a}_\omega^K}^*(v_1, v_2)\big| \leq (2+\eta+\kappa+\|BK\|_{\Lscr(X)}+\omega) \|v_1\|_V \|v_2\|_V \FORALL v_1, v_2\in V,
\end{equation}
\begin{equation}\label{eq:a_Kstar2}
 {\mathfrak{a}_\omega^K}^*(v,v)+\left(\|BK\|_{\Lscr(X)}+ \omega\right)\|v\|^2_{X} \geq \min \{\eta,\kappa\}\|v\|^2_V \FORALL v\in V.
\end{equation}
Now observe from \eqref{eq:sesqA} and \eqref{eq:sesq-Astar} that
$$ \mathfrak{a}(v_1,v_2) = \mathfrak{a}^*(v_2,v_1) \FORALL v_1,v_2\in V.$$
Using this, \eqref{eq:A-approx}, \eqref{eq:sesq-AstarK} and the expression $A_{\omega,n}^K = A_n + B_n K P_n + \omega I$ we get
\begin{equation}
 {\mathfrak{a}_\omega^K}^*(v_1,v_2)=-\left\langle \left(A^K_{\omega,n}\right)^{*}v_1, v_2 \right\rangle_{X} \FORALL v_1,v_2\in V_n.
\end{equation}
Recall Assumption \ref{as:subspace}. Let $\zeta= \|BK\|_{\Lscr(X)}+\omega$. Applying \cite[Theorem 2.2]{BaIt:1988} to the bilinear form ${\mathfrak{a}_\omega^K}^*$ it follows that $\zeta$ is in the resolvent set of $A^K_{\omega}$, $(A^{K}_{\omega})^{*}$, $A^K_{\omega,n}$ and $(A^{K}_{\omega,n})^{*}$  for all $n\in\nline$ and \vspace{-1mm}
\begin{equation} \label{eq:resolest}
  \lim_{n\to \infty}\big\| \left(\zeta I-\left(A^{K}_{\omega,n}\right)^{*}\right)^{-1}P_n x - \left(\zeta I-\left(A^{K}_{\omega}\right)^{*}\right)^{-1} x\big\|_{X}=0 \FORALL x\in X. \vspace{-1mm}
\end{equation}

Next, we present an important result which will be used to prove Proposition \ref{pr:unifstab}. For a discussion on the significance of this result, see Remark \ref{rm:compactness}. \vspace{-2mm}

%%%%%%%%%%**********%%%%%%%%%%**********%%%%%%%%%%**********%%%%%%%%%%%
%-------------------------- Begin Prop ----------------------------%
\begin{framed} \vspace{-2mm}
\begin{proposition}\label{pr:approxeig}
Let $(n_k)_{k\in\nline}$ be an increasing sequence of positive integers, $(\lambda_k)_{k\in\nline}$ be a sequence of complex numbers, $(v_k)_{k\in\nline}$ and $(w_k)_{k\in\nline}$ be two sequences in $X$ with $\|v_k\|_X=1$ and $v_k, w_k \in V_{n_k}$ for each $k\in\nline$. Recall $\omega$, $\omega_0$, $A^K_{\omega}$ and $A^K_{\omega,n}$ from \eqref{eq:lenovo} and the text above it. Suppose that
\begin{align}
 &A_{\omega,{n_k}}^K v_k = \lambda_k v_k + w_k \FORALL k\in\nline, \label{eq:assump1}\\
 &\lim_{k\to\infty}\|w_k\|_X=0, \qquad \lim_{k\to\infty} \lambda_k=\lambda \label{eq:assump2}
\end{align}
for some $\lambda\in\cline$ with $\lambda\neq \omega-\omega_0$. Then there exists a non-zero $v\in \Dscr(A)$ and a subsequence $(v_{k_r})_{r\in \nline}$ of $(v_k)_{k\in \nline}$ such that \vspace{-1mm}
\begin{equation}\label{eq:weakconv}
  \lim_{r\to\infty}\langle v_{k_r},x \rangle_{X}=\langle v,x \rangle_{X} \FORALL x\in X \vspace{-1mm}
\end{equation}
and $\lambda$ is an eigenvalue of $A_{\omega}^K$ with eigenvector $v$, i.e. $A_{\omega}^K v =\lambda v$. \vspace{-1mm}
\end{proposition}
\end{framed}
%--------------------------- End Prop -----------------------------%
%%%%%%%%%%**********%%%%%%%%%%**********%%%%%%%%%%**********%%%%%%%%%%%

%%%%%%%%%%%%%%%%%%%%%%%%%%%%-- Proof --%%%%%%%%%%%%%%%%%%%%%%%%%%%%%
\begin{proof}
Suppose that
$$ v_k=\bbm{\phi_k \\ \psi_k}, \qquad w_k=\bbm{p_k \\ q_k} $$
so that $\phi_k, \psi_k, p_k, q_k\in H^1_0(\Omega)$. Using this notation, the assumption $v_k\in V_{n_k}$ and the expression $A^K_{\omega,n_k}=A_{n_k}+P_{n_k}BKP_{n_k}+\omega I$ we can rewrite \eqref{eq:assump1} \vspace{-1mm} as
\begin{equation}\label{eq:mastereq}
 -A_{n_k}\bbm{\phi_k \\ \psi_k}=P_{n_k}BK\bbm{\phi_k \\ \psi_k}-(\lambda_k-\omega)\bbm{\phi_k \\ \psi_k}-\bbm{p_k \\ q_k}. \vspace{-1mm}
\end{equation}
Taking the inner product of the above equation with $\sbm{\phi_k \\ \psi_k}\in V_{n_k}$ in $X$ and then using \eqref{eq:A-approx} and the assumption $\big\|\sbm{\phi_k \\ \psi_k}\big\|_{X}=1$ we get
\begin{equation} \label{eq:masterinp}
 \mathfrak{a}\left(\bbm{\phi_k \\ \psi_k},\bbm{\phi_k \\ \psi_k}\right)=\left\langle BK\bbm{\phi_k \\ \psi_k},\bbm{\phi_k \\ \psi_k}\right\rangle_{X} +\left(\omega-\lambda_k\right) -\left\langle \bbm{p_k \\ q_k},\bbm{\phi_k \\ \psi_k} \right\rangle_{X}.
\end{equation}
Using \eqref{eq:sesqAcoer} to lower bound the term on the left-side of \eqref{eq:masterinp} and using the Cauchy-Schwarz inequality to upper bound the terms on right-side of \eqref{eq:masterinp} we get
$$\left\|\bbm{\phi_k \\ \psi_k}\right\|^2_{V} \leq \frac{1}{\min\{\eta,\kappa\}} \left(\|BK\|_{\Lscr(X)}+\omega+|\lambda_k|+\left\|\bbm{p_k \\ q_k}\right\|_{X}\right). $$
From \eqref{eq:assump2} it follows that the terms on right-side of the above equation can be bounded by a constant independent of $k$ and therefore the same is true for the term on the left-side, i.e. the sequences $(\phi_k)_{k\in \nline}$ and $(\psi_k)_{k\in \nline}$ are uniformly bounded in $H^1_0(\Omega)$. Using this and the fact that $H^1_0(\Omega)$ is compactly embedded in $L^2(\Omega)$ we can conclude that there exist limits $\phi,\psi\in H^1_0(\Omega)$ and subsequences of $(\phi_k)_{k\in \nline}$ and $(\psi_k)_{k\in \nline}$ which converge, weakly in $H^1_0(\Omega)$ and strongly in $L^2(\Omega)$, to $\phi$ and $\psi$, respectively. More specifically, there exists an increasing sequence of positive integers $(k_r)_{r\in\nline}$ such that
\begin{equation}\label{eq:weakphipsi}
\lim_{r\to \infty} \langle \phi_{k_r}-\phi,z \rangle_{H^1_0(\Omega)}=0, \qquad
\lim_{r\to \infty} \langle \psi_{k_r}-\psi,z \rangle_{H^1_0(\Omega)}=0 \FORALL z\in H^1_0(\Omega),
\end{equation}
\begin{equation}\label{eq:strongphipsi}
  \lim_{r \to \infty} \|\phi_{k_r} - \phi\|_{L^2(\Omega)} = 0, \qquad \lim_{r \to \infty} \|\psi_{k_r} - \psi\|_{L^2(\Omega)} = 0.
\end{equation}
We remark that to arrive at the above conclusions we have used the facts that a weakly convergent sequence in $H_0^1(\Omega)$ is also weakly convergent in $L^2(\Omega)$ with the same limit, and the weak and strong limits of a sequence in $L^2(\Omega)$ (whenever they exist) are the same.

Define $v=\sbm{\phi \\ \psi}$ and recall that $v_{k_r}=\sbm{\phi_{k_r}\\ \psi_{k_r}}$. It follows directly from \eqref{eq:weakphipsi}-\eqref{eq:strongphipsi} that \eqref{eq:weakconv} holds. We will now show that \vspace{-0.5mm}
\begin{equation}\label{eq:vprop}
 v\in \Dscr(A), \qquad A^K_{\omega}v=\lambda v. \vspace{-0.5mm}
\end{equation}
Let $\zeta= \|BK\|_{\Lscr(X)}+\omega$. Using \eqref{eq:assump1} it is easy to see that \vspace{-1mm}
$$\big(\zeta-A^{K}_{\omega,n_{k_r}}\big)v_{k_r}= (\zeta-\lambda_{k_r})v_{k_r}-w_{k_r}. \vspace{-1mm}$$
The resolvent set of $A^{K}_{\omega,n_{k_r}}$ contains $\zeta$ (see the discussion above \eqref{eq:resolest}) and so the above equation can be rewritten as \vspace{-0.5mm}
$$v_{k_r}=\big(\zeta I -A^{K}_{\omega,n_{k_r}}\big)^{-1} \big[(\zeta -\lambda_{k_r}) v_{k_r}-w_{k_r}\big]. \vspace{-0.5mm}$$
Taking the inner product of this equation in $X$ with elements belonging to $V_{n_{k_r}}=P_{n_{k_r}}X$ and then using $((\zeta I -A^K_{\omega,n_{k_r}})^{-1})^{*} = (\zeta I - (A^K_{\omega,n_{k_r}})^{*})^{-1}$ gives \vspace{-0.5mm}
\begin{equation}\label{eq:resolvlim}
 \langle v_{k_r},P_{n_{k_r}} x \rangle_{X}=\big \langle \left[(\zeta -\lambda_{k_r})v_{k_r}-w_{k_r}\right], \big(\zeta I - \big(A^K_{\omega,n_{k_r}}\big)^{*}\big)^{-1} P_{n_{k_r}} x \big \rangle_{X} \ \quad\forall x\in X.\vspace{-0mm}
\end{equation}
From \eqref{eq:assumconv} and \eqref{eq:weakconv} (which we have established above) we get \vspace{-1mm}
$$ \lim_{r\to\infty}\left\|P_{n_{k_r}} x - x\right\|_X = 0,\quad \qquad \lim_{r\to\infty}\langle v_{{k_r}}, x\rangle_X = \langle v, x\rangle_X , \vspace{-1mm}$$
using which it follows that $\lim_{r\to \infty}\langle v_{k_r}, P_{n_{k_r}} x \rangle_{X} = \langle v, x \rangle_{X}$. Similarly, from \eqref{eq:resolest}, \eqref{eq:assump2} and \eqref{eq:weakconv} we get \vspace{-1mm}
$$ \lim_{r\to\infty} \left\|\big(\zeta I- \big(A^K_{\omega,n_{k_r}}\big)^{*}\big)^{-1} P_{n_{k_r}}x - \big(\zeta I - \big(A^K_{\omega}\big)^{*}\big)^{-1} x\right\|_X = 0, $$
$$ \lim_{r\to\infty}\langle(\zeta-\lambda_{k_r})v_{k_r}-w_{n_r}, x\rangle_X = \langle (\zeta-\lambda)v,x\rangle_X ,\vspace{-0.5mm}$$
using which it follows that the term on the right-side of \eqref{eq:resolvlim} converges to the expression $\langle(\zeta-\lambda) v, (\zeta I-(A_\omega^K)^*)^{-1} x\rangle_X$ as $r\to \infty$. Consequently, taking the limit as $r\to \infty$ in \eqref{eq:resolvlim} and then using $((\zeta I - (A^K_{\omega})^{*})^{-1})^{*}=(\zeta I -A^K_{\omega})^{-1}$  gives \vspace{-0.5mm}
$$ \langle v , x \rangle_{X} = \left \langle (\zeta-\lambda)(\zeta I -A^K_{\omega})^{-1}v, x \right \rangle_{X} \FORALL x\in X. \vspace{-0.5mm}$$
It follows from the above equation that $v=(\zeta-\lambda)(\zeta I -A^K_{\omega})^{-1}v$ which implies that $v\in \Dscr(A)$ and (after a simple calculation) that $A^K_\omega v=\lambda v$, i.e. \eqref{eq:vprop} holds.

We will now complete the proof of this proposition by showing that $v\neq 0$, which will imply
that $\lambda$ is indeed an eigenvalue of $A^K_\omega$ with corresponding eigenvector $v$. To this end we suppose that \vspace{-1mm}
\begin{equation} \label{eq:opposite}
 \phi=0, \qquad \psi=0. \vspace{-1mm}
\end{equation}
We will show below that this leads to a contradiction and so $v=\sbm{\phi \\ \psi}$ cannot be 0.

Since $\sbm{\phi_{k_r}\\ \psi_{k_r}}\in V_{n_{k_r}}$, it follows that $\sbm{\psi_{k_r}\\ \eta \psi_{k_r}}\in V_{n_{k_r}}$. Letting $k=k_r$ in \eqref{eq:mastereq} and then taking its inner product with $\sbm{\psi_{k_r}\\ \eta \psi_{k_r}}$ in $X$ and subsequently using \eqref{eq:A-approx} we get
\begin{align}
 \mathfrak{a}\left(\bbm{\phi_{k_r} \\ \psi_{k_r}},\bbm{\psi_{k_r} \\ \eta \psi_{k_r}}\right) = &\left\langle BK \bbm{\phi_{k_r} \\ \psi_{k_r}},\bbm{\psi_{k_r} \\ \eta \psi_{k_r}} \right\rangle_{X} +\left(\omega-\lambda_{k_r}\right) \left\langle \bbm{\phi_{k_r}\\ \psi_{k_r}}, \bbm{\psi_{k_r} \\ \eta \psi_{k_r}} \right\rangle_{X} \nonumber\\[1ex]
 &\hspace{10mm} -\left\langle \bbm{p_{k_r} \\ q_{k_r}},\bbm{\psi_{k_r} \\ \eta \psi_{k_r}} \right\rangle_{X}. \label{eq:mistakes}\\[-5ex]\nonumber
\end{align}
Replacing the term on the left-side of \eqref{eq:mistakes} using the definition of the bilinear form $\mathfrak{a}$ in \eqref{eq:sesqA}, rewriting the first term on the right-side using $\langle B \alpha, \sbm{y\\z}\rangle_{X}=\langle B \alpha, \sbm{y\\0}\rangle_{X}$  which follows from the definition of $B$ in \eqref{eq:Bdef}, 
expressing the second term on the right-side of \eqref{eq:mistakes} in terms of the inner products in $L^2(\Omega)$ and $H^1_0(\Omega)$ and finally rearranging the resulting terms we get \vspace{-1mm}
\begin{align}
    \mu_{r}\|\psi_{k_r}\|^2_{H^1_0(\Omega)} =&\left\langle BK \bbm{\phi_{k_r} \\ \psi_{k_r}},\bbm{\psi_{k_r} \\ 0} \right\rangle_{X} -\left\langle \bbm{p_{k_r} \\ q_{k_r}}, \bbm{\psi_{k_r}\\ \eta \psi_{k_r}} \right\rangle_{X} \nonumber \\[1ex]
    &\hspace{10mm}+\left(\omega-\lambda_{k_r}\right)\langle \phi_{k_r},\psi_{k_r} \rangle_{L^2(\Omega)}, \label{eq:forestimate}
\end{align}
where $\mu_{r}=\left(1+\eta (\kappa-\omega+\lambda_{k_r})\right)$. From \eqref{eq:assump2} and the definition of $\omega_0$ we get that $\lim_{r\to\infty}\mu_r= \eta(\omega_0 - \omega+\lambda)$. Since $\eta>0$ and $\lambda\neq \omega -\omega_0$ by assumption, it follows that \vspace{-1mm}
\begin{equation} \label{eq:alooparatha}
 \lim_{r\to\infty}\mu_r \neq0. \vspace{-1mm}
\end{equation}
Using the Cauchy–Schwarz inequality and the estimate $\left\|\sbm{\phi_{k_r} \\ \psi_{k_r}}\right\|_{X}=1$, the first term on the right-side of \eqref{eq:forestimate} can be bounded by
$\|BK\|_{\Lscr(X)}\|\psi_{k_r}\|_{L^2(\Omega)}$,
which converges to zero as $r\to\infty$, see \eqref{eq:strongphipsi} and \eqref{eq:opposite}. Hence \vspace{-1mm}
\begin{equation}\label{eq:idly}
\lim_{r\to \infty}\left| \left\langle  BK\bbm{\phi_{k_r}\\ \psi_{k_r}}, \bbm{\psi_{k_r}\\ 0} \right\rangle_{X} \right|=0. \vspace{-1mm}
\end{equation}
Since $(\psi_{k_r})_{r\in \nline}$ is uniformly bounded in $H^1_0(\Omega)$, see discussion above \eqref{eq:weakphipsi}, it follows that $\left(\left\|\sbm{\psi_{k_r}\\ \eta \psi_{k_r}}\right\|_{X}\right)_{r\in \nline}$ is uniformly bounded in $X$. This and the first limit in \eqref{eq:assump2} 
imply that \vspace{-1mm}
\begin{equation}\label{eq:upma}
\lim_{r\to \infty}\left| \left\langle \bbm{p_{k_r} \\ q_{k_r}}, \bbm{\psi_{k_r}\\ \eta \psi_{k_r}} \right\rangle_{X} \right|=0. \vspace{-1mm}
\end{equation}
Finally, it follows from \eqref{eq:assump2}, \eqref{eq:strongphipsi} and \eqref{eq:opposite} that \vspace{-1mm}
\begin{equation}  \label{eq:sprouts} \lim_{r\to\infty}\left(\omega-\lambda_{k_r}\right)\langle \phi_{k_r},\psi_{k_r} \rangle_{L^2(\Omega)}=0. \vspace{-1mm}
\end{equation}
Using the limits in \eqref{eq:alooparatha}-\eqref{eq:sprouts}, we can conclude from \eqref{eq:forestimate} that $\lim_{r\to\infty} \|\psi_{k_r}\|_{H_0^1(\Omega)}=0$. From \eqref{eq:strongphipsi} and \eqref{eq:opposite} we have $\lim_{r\to\infty} \|\phi_{k_r}\|_{L^2(\Omega)}=0$. Combining these we get \vspace{-2mm}
$$ \lim_{r\to\infty} \left\| \bbm{\phi_{k_r} \\ \psi_{k_r}}\right\|_X =0, \vspace{-2mm}$$
which contradicts the assumption in the proposition that $\|v_{k_r}\|_X=1$ for all $r\in\nline$. So \eqref{eq:opposite} cannot hold and $v=\sbm{\phi \\ \psi}$ must be a nontrivial element in $X$.

In summary, we have shown that there exists a non-zero $v\in X$ such that \eqref{eq:weakconv} and \eqref{eq:vprop} hold. This completes the proof of the proposition. \vspace{-1mm}
\end{proof}
%%%%%%%%%%%%%%%%%%%%%%%%%-- End Proof --%%%%%%%%%%%%%%%%%%%%%%%%%%%%

In the next proposition, we establish the uniform (in $n$) stabilizability of the pair $(A_{\omega,n},B_n)$. In particular, we show that for each $n\in\nline$ sufficiently large there exists a $K_n\in\Lscr(V_n,\rline^m)$ such that $\|e^{(A_{\omega,n}+B_n K_n)t} x\|_X \leq M e^{-\nu t} \|x\|_X$ for all $x\in V_n$, each $t\geq0$ and some $M,\nu>0$ independent of $n$.

%%%%%%%%%%%%%%%%%%%%%%%%-- Begin Theorem -- %%%%%%%%%%%%%%%%%%%%%%%%
\begin{framed} \vspace{-2mm}
\begin{proposition} \label{pr:unifstab}
Let $A^K_{\omega,n}=A_{\omega,n}+B_nKP_n$ be as defined in \eqref{eq:lenovo}. There exists an $n_0\in \nline$ such that for each $n>n_0$ we have \vspace{-1mm}
\begin{equation}\label{eq:exponential-decay}
  \|e^{A^K_{\omega,n} t} x\|_X \leq Me^{-\nu t}\|x\|_X \FORALL x\in V_n,\quad \forall t>0 \vspace{-1mm}
\end{equation}
and for some $M,\nu>0$ independent of $n$. \vspace{-2mm}
\end{proposition}
\end{framed}
%%%%%%%%%%%%%%%%%%%%%%%%-- End Theorem -- %%%%%%%%%%%%%%%%%%%%%%%%
\vspace{-6mm}

%%%%%%%%%%%%%%%%%%%%%%%%%-- Begin Proof -- %%%%%%%%%%%%%%%%%%%%%%%%
\begin{proof}
We will complete the proof of this proposition in two steps. In the first step we will establish that there exist an integer $n_0>0$ and constants $\nu>0$ and $\theta\in(\frac{\pi}{2},\pi)$ such that the spectrum of $A^K_{\omega,n}$ lies in the open sector \vspace{-1mm}
\begin{equation}\label{eq:sectornu}
  \Sigma = \big\{\lambda\in\cline \,\big|\, |\arg(\lambda+\nu)|>\theta\big\} \vspace{-1mm}
\end{equation}
for all $n>n_0$. Here $\arg$ takes values in $(-\pi,\pi]$. In the second step we will show that the resolvent $(\lambda I - A^K_{\omega,n})^{-1}$ is uniformly bounded on the boundary of the sector $\Sigma$ with the bound being uniform in $n>n_0$. The estimate in \eqref{eq:exponential-decay} will then follow directly from the integral representation of semigroups in terms of the resolvent.

\noindent
\textbf{Step 1.} Consider the bilinear form $\mathfrak{a}_{\omega, n}^K: V_n \times V_n \rightarrow \mathbb{R}$ defined as
\begin{equation}\label{eq:sesq-a-K}
 \mathfrak{a}_{\omega, n}^K(v_1,v_2)= \mathfrak{a} \left(v_1,v_2\right) -\langle BK v_1, v_2 \rangle_X - \omega \langle v_1, v_2 \rangle_X \FORALL v_1, v_2\in V_n.
\end{equation}
From \eqref{eq:A-approx} it follows that the operator associated with $\mathfrak{a}_{\omega, n}^K$ is $A^K_{\omega,n}$, i.e.
\begin{equation}
 {\mathfrak{a}_{\omega,n}^K}(v_1,v_2)=-\left\langle A^K_{\omega,n} v_1, v_2 \right\rangle_{X} \FORALL v_1,v_2\in V_n.
\end{equation}
Let $\zeta=\|BK\|_{\Lscr(X)}+\omega$. Using the estimates in \eqref{eq:sesqAcont} and \eqref{eq:sesqAcoer} we get the following continuity and coercivity estimates for $\mathfrak{a}_{\omega, n}^K$: \vspace{-1mm}
\begin{equation}\label{eq:a_K one}
 |\mathfrak{a}_{\omega, n}^K(v_1,v_2)| \leq (2+\eta +\kappa +\zeta)\|v_1\|_V \|v_2\|_V \FORALL v_1,v_2\in V_n,
\end{equation}
\begin{equation}\label{eq:a_K two}
 \mathfrak{a}^K_{\omega,n}(v,v)+\zeta\|v\|^2_X \geq \min\{\eta,\kappa\}\|v\|^2_V \FORALL v\in V_n.
\end{equation}
Applying \cite[Chapter IV, Theorem 6.1]{Sh:2010} to $\mathfrak{a}^K_{\omega,n}$ we can conclude that there exist constants $\theta_0 \in\left(\frac{\pi}{2}, \frac{3\pi}{4}\right)$ and $M_0>0$ which are independent of $n$ (they depend only on the constants in \eqref{eq:a_K one}-\eqref{eq:a_K two}) such that the spectrum of $A^K_{\omega,n}$ is contained in the open sector \vspace{-1mm}
\begin{equation} \label{eq:sigma0}
 \Sigma_0 = \left\{\lambda \in \cline \ \big| \ |\arg(\lambda -\zeta)|> \theta_0  \right\} \vspace{-1mm}
\end{equation}
for all $n\in\nline$, \vspace{-2mm}
\begin{equation}\label{eq:resestout}
 \sup_{v\in V_n,\ \|v\|_X=1}\big\|\big(\lambda I-A_{\omega, n}^K \big)^{-1} v \big\|_X \leq \frac{M_0}{\left|\lambda-\zeta\right|} \FORALL \lambda\in \cline\setminus \overline{\Sigma_0} \vspace{-1mm}
\end{equation}
and all $n\in\nline$ and \vspace{-1mm}
\begin{equation} \label{eq:contrest}
 \|e^{A^K_{\omega,n} t}x\|_X \leq e^{\zeta t}\|x\|_X \FORALL x\in V_n, \quad \forall t>0, \quad \forall n\in\nline. \vspace{-1mm}
\end{equation}
Here $\overline{\Sigma_0}$ denotes the closure of the set $\Sigma_0$. Let $A^K_{\omega}=A_{\omega}+BK$ be the exponentially stable operator defined in \eqref{eq:lenovo}. Recall that $\omega$ is chosen so that $0<\omega<\omega_0$. Fix $\beta\in(0,\omega_0-\omega)$ such that the spectrum of $A^K_{\omega}$ is contained in the left half-plane $\cline^-_{-\beta}$. Then there exists an integer $n_0>0$ such that the spectrum of $A^K_{\omega,n}$ is also contained in $\cline^-_{-\beta}$ for all $n>n_0$. Indeed, if this were not true, then there would exist an increasing sequence of positive integers $(n_k)_{k\in\nline}$ and a converging sequence of complex numbers $(\lambda_k)_{k\in\nline}$ in $\overline{\cline^+_{-\beta}}\cap\Sigma_0$ such that $\lambda_k$ is an eigenvalue of $A^K_{\omega,n_k}$ for each $k\in\nline$. This would then imply via Proposition \ref{pr:approxeig} that the limit $\lambda\in\overline{\cline^+_{-\beta}}$ of the sequence $(\lambda_k)_{k\in\nline}$ is an eigenvalue of $A^K_{\omega}$, which is a contradiction. So the spectrum of $A^K_{\omega,n}$ is contained in $\cline^-_{-\beta}$ for all $n>n_0$ and it is also contained in $\Sigma_0$ for all $n\in\nline$ (see the discussion above \eqref{eq:sigma0}). Therefore there exist $\nu>0$ and $\theta\in(\frac{\pi}{2},\pi)$ such that the spectrum of $A^K_{\omega,n}$ lies in the sector $\Sigma$ defined in \eqref{eq:sectornu} with $\Sigma \cap \Sigma_0\subset \overline{\cline^+_{-\beta}}$, see Figure 1. \vspace{4mm}

%%%%%%%%%%%%%%%%%%%%%%%%%%%%% Figure 1 %%%%%%%%%%%%%%%%%%%%%%%%%%%%%
\begin{figure}[htp!]
    \centering
    \includegraphics[scale=0.6]{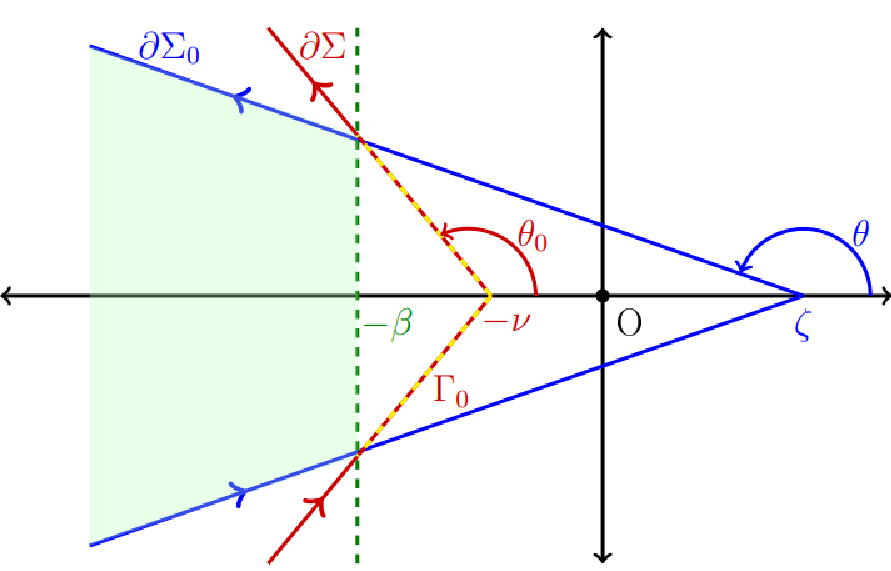}
    \parbox{5.2in}{\caption{\small{The boundaries of the sectors $\Sigma$ in \eqref{eq:sectornu} and $\Sigma_0$ in \eqref{eq:sigma0} are shown using red and blue colored rays, respectively. The yellow dashed line represents the boundary of $\Sigma$ which lies inside $\overline{\Sigma_0}$. In Step 1 of the proof of Proposition \ref{pr:unifstab} we have shown that (for $n$ sufficiently large) the eigenvalues of $A^K_{\omega,n}$ are contained in $\Sigma_0$ and have real part less than $-\beta$, i.e. they lie in the green shaded region. Hence they are contained in $\Sigma$.} }}
    \label{fig:Contours}
\end{figure}
%%%%%%%%%%%%%%%%%%%%%%%%%%%%% Figure 1 %%%%%%%%%%%%%%%%%%%%%%%%%%%%%

\noindent
\textbf{Step 2.} From \eqref{eq:resestout} we get that the resolvent operator $(\lambda I - A^K_{\omega,n})^{-1}:V_n\to V_n$ is uniformly bounded (in the $\Lscr(X)$ norm) on the boundary of the sector $\Sigma$ which lies in $\cline\setminus\overline{\Sigma_0}$ with the bound being uniform in $n$. We claim that it is also uniformly bounded on the compact set $\Gamma_0$, which is the boundary of $\Sigma$ which lies in $\overline{\Sigma_0}$, with the bound being uniform in $n>n_0$. Indeed, if this were not true, then there would exist an increasing sequence of integers $(n_k)_{k\in\nline}$ with $n_1>n_0$, a converging sequence of complex numbers $(\lambda_k)_{k\in\nline}$ in $\Gamma_0$ and two sequences $(v_k)_{k\in\nline}$ and $(w_k)_{k\in\nline}$ in $X$ with $v_k, w_k\in V_{n_k}$ and $\|v_k\|_X=1$ for each $k\in\nline$ such that $\lim_{k\to\infty}\|w_k\|_X=0$ and \vspace{-1mm}
$$ (\lambda_k I - A_{\omega,n_k}^K)^{-1} w_k =v_k. \vspace{-1mm} $$
This would then imply via Proposition \ref{pr:approxeig} that the limit $\lambda\in\Gamma_0\subset\overline{\cline^+_{-\beta}}$ of the sequence $(\lambda_k)_{k\in\nline}$ is an eigenvalue of $A^K_{\omega}$, which is a contradiction. So $(\lambda I - A^K_{\omega,n})^{-1}$ is uniformly bounded on the boundary of the sector $\Sigma$, i.e. there exists $M_1>0$ independent of $n$ such that \vspace{-1mm}
\begin{equation}\label{eq:resolsig}
 \sup_{v\in V_n,\ \|v\|_X=1}\big\|\big(\lambda I-A_{\omega, n}^K \big)^{-1} v \big\|_X \leq M_1 \FORALL \lambda\in \partial\Sigma \vspace{-1.5mm}
\end{equation}
and all $n>n_0$. Here $\partial\Sigma$ denotes the boundary of $\Sigma$

We will now complete the proof of this proposition. Recall that $A^{K}_{\omega,n}\in \Lscr(V_n)$. % and $V_n$ is a finite-dimensional space.
For each $n>n_0$, from the Dunford integral formula 
and the conclusion in Step 1 that the spectrum of $A^K_{\omega,n}$ is contained in $\Sigma$, we \vspace{-1mm} get that
$$ e^{A^{K}_{\omega,n}t}x = \frac{1}{2\pi i}\int_{\partial\Sigma} e^{\lambda t} (\lambda I-A^K_{\omega,n})^{-1}x\, \dd \lambda \FORALL x\in V_n, \quad\forall t>0. \vspace{-1mm} $$
Note that the above integral is a contour integral that must be computed anti-clockwise along the boundary $\partial\Sigma$ of $\Sigma$. The boundary $\partial\Sigma$ consists of two rays: $\lambda=-\nu+r e^{-i\theta}$ and  $\lambda=-\nu+r e^{i\theta}$ with $r\in[0,\infty)$, see Figure 1. Rewriting the above integral as a sum of two integrals, one along each ray, and then changing the integration variable from $\lambda$ to $r$ and finally using \eqref{eq:resolsig} to bound the two integrands, it follows after a simple calculation that for each $n>n_0$, \vspace{-1mm}
$$ \|e^{A^{K}_{\omega,n}t}x\|_X \leq \frac{M_1}{\pi}\left( \frac{e^{-\nu t}}{t|\cos\theta|} \right)\|x\|_{X}\FORALL x\in V_n, \quad\forall t>0. \vspace{-1mm} $$
Using the estimate in \eqref{eq:contrest} for $t\in[0,1]$ and the above estimate for $t>1$, it follows that \eqref{eq:exponential-decay} holds for each $n>n_0$ with $M=\max\{e^{\zeta+\nu},M_1/(\pi|\cos\theta|)\}$.
\end{proof}
%%%%%%%%%%%%%%%%%%%%%%%%%%%-- End Proof -- %%%%%%%%%%%%%%%%%%%%%%%%

The next remark discusses the connections (and differences) between the results in \cite{BaKu:1984} and this work. In particular, it highlights the significance of \vspace{1mm} Proposition \ref{pr:approxeig}.

%%%%%%%%%%%%%%%%%%%%%%%%%-- Begin remark -- %%%%%%%%%%%%%%%%%%%%%%%%
\begin{remark} \label{rm:compactness}
In this work, we have addressed the $\omega$-stabilization problem for a heat equation with memory by adapting techniques developed for parabolic PDEs in \cite{BaKu:1984}, see the proof of our main result Theorem \ref{thm:main_result}. Central to applying the techniques from \cite{BaKu:1984} to the heat equation with memory is establishing the uniform (in $n$) stabilizability of the finite-dimensional approximation \eqref{eq:approx system} associated with the abstract evolution equation \eqref{eq:HM_absomega}; We have established this uniform stabilizability result in Proposition \ref{pr:unifstab}. The proof of this proposition depends substantially on the conclusions of Proposition \ref{pr:approxeig}, which we have in turn established by combining arguments from \cite{BaKu:1984} with certain novel arguments (that were not needed in \cite{BaKu:1984}). More specifically, our proof of Proposition \ref{pr:approxeig} consists of three steps: The  first step shows that the sequence $(v_k)_{k\in\nline}$ is bounded in $V$ and has a subsequence $(v_{k_r})_{r\in\nline}$ which converges weakly in $X$ to $v$ (see discussion above \eqref{eq:vprop}). The second step establishes that $v \in D(A)$ and $A_\omega^K v=\lambda v$ (see discussion below \eqref{eq:vprop}). The third step shows that $v \neq 0$, which confirms that $\lambda$ is indeed an eigenvalue of $A_\omega^K$ (see discussion below \eqref{eq:opposite}). The arguments in the first two steps are direct adaptations of arguments from \cite[Lemma 3.3]{BaKu:1984} to the heat equation with memory. In the case of the parabolic PDEs considered in \cite{BaKu:1984}, the claim $v\neq0$ established in our third step holds trivially. Indeed, $V=H^1_0(\Omega)$ and $X=L^2(\Omega)$ in \cite{BaKu:1984} so that $V$ is compactly embedded in $X$. Consequently, the sequence $(v_k)_{k\in\nline}$ which is bounded in $V$ and satisfies $\|v_k\|_X=1$ has a subsequence $(v_{k_r})_{r\in\nline}$ which converges strongly in $X$ to $v$ with $\|v\|_X=1$ (i.e. $v\neq0$). However, in the case of the heat equation with memory considered in this work $V=H^1_0(\Omega)\times H^1_0(\Omega)$ is not compactly embedded in $X=L^2(\Omega)\times H^1_0(\Omega)$ and hence additional arguments are needed to show that $v\neq0$.
\end{remark}
%%%%%%%%%%%%%%%%%%%%%%%%%--  End remark  --%%%%%%%%%%%%%%%%%%%%%%%%

Next we present the main result of this paper. Recall the state operator $A$ and the constants $\kappa, \eta>0$ from \eqref{eq:Adom}-\eqref{eq:Adef}, the control operator $B$ from \eqref{eq:Bdef} and $B^*$ from \eqref{eq:Bdef-star}. From Section \ref{sec4.1} recall the subspace $V_n$, the projection operator $P_n:X\to V_n$ and the approximations $A_n$ and $B_n$ of $A$ and $B$, respectively. Finally, recall the self-adjoint coercive operator $Q$ and the positive definite matrix $R$ used to define the cost functional in \eqref{eq:Cost} and let $Q_n=P_n Q P_n$. We now state the theorem. \vspace{-1mm}

%%%%%%%%%%%%%%%%%%%%%%%%%-- Begin Theorem -- %%%%%%%%%%%%%%%%%%%%%%%
\begin{framed} \vspace{-2mm}
\begin{theorem} \label{thm:main_result}
Let $\omega_0=\kappa+\frac{1}{\eta}$. Fix $0<\omega<\omega_0$. Define $A_\omega = A + \omega I$ and $A_{\omega,n} = A_n + \omega I$. Let Assumption \ref{as:subspace} hold. Suppose that $B$ satisfies the hypothesis in Theorem \ref{thm:ABstab}. Then there exists a unique nonnegative solution $\Pi\in\Lscr(X)$ to the operator Riccati equation \eqref{eq:Riccati} and a unique nonnegative solution $\Pi_n\in \Lscr(V_n)$ to the finite-dimensional Riccati equation \eqref{eq:Riccati_approx} for each $n>n_0$ (here $n_0>0$ is the integer in Proposition \ref{pr:unifstab}) such that \vspace{-2.5mm}
\begin{equation}\label{eq:Ricc-sol-conv}
  \lim_{\substack{n\to \infty\\ n>n_0}}\|\Pi_nP_nx-\Pi x\|_{X}=0 \FORALL x\in X. \vspace{-3mm}
\end{equation}
The feedback gain $K_\infty=-R^{-1}B^{*}\Pi$ stabilizes \eqref{eq:HM_absomega} and the feedback gain $K_n=-R^{-1}B^{*}_n\Pi_n$ stabilizes \eqref{eq:approx system} for each $n>n_0$ and \vspace{-2.5mm}
\begin{equation}\label{eq:Con-gain-conv}
  \lim_{\substack{n\to \infty\\ n>n_0}} \|K_n P_n-K_\infty\|_{\Lscr(X,\rline^m)}=0. \vspace{-3mm}
\end{equation}
In particular, the controller gain $K_\omega=K_n P_n$ with $n$ sufficiently large solves the $\omega$-stabilization problem, Problem \ref{prob:wstab}.
\end{theorem} \vspace{-2mm}
\end{framed}
%%%%%%%%%%%%%%%%%%%%%%%%%%-- End Theorem -- %%%%%%%%%%%%%%%%%%%%%%%%
\vspace{-4mm}
%%%%%%%%%%%%%%%%%%%%%%%%%%-- Begin Proof -- %%%%%%%%%%%%%%%%%%%%%%%%
\begin{proof}
Under the assumptions of this theorem, there exists a unique nonnegative solution $\Pi\in\Lscr(X)$ to the operator Riccati equation \eqref{eq:Riccati} and $A_\omega-BR^{-1} B^*\Pi$ generates an exponentially stable semigroup, see the discussion below \eqref{eq:Riccati}. From Proposition \ref{pr:unifstab} we have that $A_{\omega,n}+B_nKP_n$ generates an exponentially stable semigroup for $n>n_0$, i.e. the pair $(A_{\omega,n},B_n)$ is stabilizable. Furthermore, $Q_n$ is a coercive operator. It now follows from \cite[Theorem 6.2.7]{CuZw:1995} that there exists a unique nonnegative solution $\Pi_n\in\Lscr(V_n)$ to the finite-dimensional Riccati equation \eqref{eq:Riccati_approx} for each $n>n_0$. The result \cite[Theorem 2.2]{BaKu:1984} ensures that the limit in \eqref{eq:Ricc-sol-conv} holds provided the following five conditions hold for $n>n_0$: \vspace{-2mm}
\begin{enumerate}
  \item[(C1)] For each initial state of \eqref{eq:approx system}, there exists an input $u\in L^2((0,\infty); \rline^m)$ for \eqref{eq:approx system} such that the corresponding cost \eqref{eq:Cost_approx} is finite. Furthermore, whenever the cost \eqref{eq:Cost_approx} associated with an initial state and input is finite, then the corresponding state trajectory of \eqref{eq:approx system} converges to zero asymptotically. \vspace{-2mm}
  \item[(C2)] The semigroup $\tline_\omega$ generated by $A+\omega I$ and the adjoint semigroup $\tline_\omega^*$ satisfy \vspace{-1mm}
      $$ \lim_{n\to \infty}\left\| \tline_\omega(t) x - e^{A_{\omega,n} t} P_n x \right\|_X=0 \FORALL x\in X, \vspace{-2mm}$$
      $$ \lim_{n\to \infty} \left\|\tline^{*}_\omega(t)x - e^{A_{\omega,n}^{*}t} P_n x \right\|_X=0 \FORALL x\in X, \vspace{-1mm}$$
      with the above convergences being uniform in $t$ on bounded subsets of \vspace{-2mm}  $[0,\infty)$.
  \item[(C3)] The operators $B$ and $B_n$ and their adjoints satisfy \vspace{-1mm}
     $$ \lim_{n\to \infty} \|B_n\alpha - B\alpha\|_X=0, \qquad \lim_{n\to \infty} \|B^{*}_n P_n x - B^{*}x\|_{\rline^m}=0 \vspace{-1mm} $$
      for all $\alpha\in \rline^m$ and $x\in X$. \vspace{-2mm}
  \item[(C4)] The operators $Q$ and $Q_n$ satisfy
      $$ \lim_{n\to \infty} \|Q_nP_n x - Qx\|_X=0 \FORALL x\in X. \vspace{-2mm}$$
  \item[(C5)] There exist positive constants $M_1, M_2$ and $\mu$ independent of $n$ such that the operators $\Pi_n$  and  $K_n=-R^{-1}B_n^{*}\Pi_n$ satisfy
       $$ \|\Pi_n x\|_X \leq M_1 \|x\|_{X} \FORALL x\in V_n, \quad \forall\, n>n_0 , $$
       $$ \left\|e^{(A_{\omega,n}+B_n K_n)t} x \right\|_X \leq M_2 e^{-\mu t}\|x\|_X \FORALL x\in V_n, \quad \forall\, n>n_0. \vspace{-2mm} $$
\end{enumerate}
We can verify that the above five conditions hold by applying the same arguments used in \cite{BaKu:1984} to verify these conditions for parabolic PDEs. We briefly summarize these arguments below. Condition (C1) holds since $(A_{\omega,n},B_n)$ is stabilizable for $n>n_0$ (see beginning of this proof) and $Q_n$ in \eqref{eq:Cost_approx} is a coercive operator. Mimicking the arguments used to establish \eqref{eq:resolest}, we can show that
$$ \lim_{n\to \infty}\big\| \left(I - A_n \right)^{-1} P_n x - \left(I - A \right)^{-1} x \big\|_X=0 \FORALL x\in X, \vspace{-2mm} $$
$$ \lim_{n\to \infty}\big\| \left(I - A_n^{*} \right)^{-1} P_n x - \left(I - A^* \right)^{-1} x \big\|_X=0 \FORALL x\in X. $$
Using these resolvent convergences and the coercivity of the bilinear forms $\mathfrak{a}$ and $\mathfrak{a}^*$, see \eqref{eq:sesqAcoer} and \eqref{eq:sesqAstarcoer}, we can conclude via the Trotter-Kato theorem \cite[Chapter 3, Theorem 4.4]{Paz:83} that Condition (C2) holds. Condition (C3) follows from the definitions of $B$ and $B_n$ and the limit in \eqref{eq:assumconv}. Note that $Q_n P_n x - Q x = P_n Q P_n x - P_n Q x + P_n Q x- Q x$ and so
$$\|Q_nP_n x - Q x\|_X \leq \|Q\|_{\Lscr(X)}\|P_n x- x\|_X + \|P_n Qx -Qx\|_{X}.$$
From this and the limit in \eqref{eq:assumconv}, Condition (C4) follows. Condition (C5) can be verified by combining the uniform stabilizability estimate \eqref{eq:exponential-decay} established in Proposition \ref{pr:unifstab} with the arguments presented after the proof of Lemma 3.3 in \cite{BaKu:1984}. Since the Conditions (C1)-(C5) hold, the limit in \eqref{eq:Ricc-sol-conv} follows from \cite[Theorem 2.2]{BaKu:1984}.

Note that $P_n\Pi_n=\Pi_n$ and $P_n B = B_n$. Taking $x=B\alpha$ in \eqref{eq:Ricc-sol-conv} we get
$$ \lim_{\substack{n\to \infty\\ n>n_0}}\|P_n\Pi_nB_n\alpha-\Pi B\alpha\|_{X}=0 \FORALL \alpha\in\rline^m. \vspace{-2mm} $$
Since $\rline^m$ is a finite-dimensional space, the above pointwise convergence implies convergence in the operator norm, i.e. $\lim_{\substack{n\to \infty\\ n>n_0}}\|P_n\Pi_nB_n-\Pi B\|_{\Lscr(\rline^m,X)}=0$ and since the norms of a bounded linear operator and its adjoint are the same we get
$$ \lim_{\substack{n\to \infty\\ n>n_0}} \|B^{*}_n\Pi_nP_n-B^{*}\Pi\|_{\Lscr(X,\rline^m)}=0. \vspace{-2mm} $$
The limit in \eqref{eq:Con-gain-conv} follows from the above limit and the definitions of $K_\infty$ and $K_n$.

Finally, since $A_\omega + B K_\infty$ generates an exponentially stable semigroup (see beginning of this proof) and \eqref{eq:Con-gain-conv} holds, it follows from the perturbation theory that $A_\omega + B K_n P_n$ also generates an exponentially stable semigroup when $n$ is sufficiently large, i.e. $K_n P_n$ with $n$ sufficiently large solves Problem \ref{prob:wstab}. This completes the proof of the theorem.
\end{proof}
%%%%%%%%%%%%%%%%%%%%%%%%%%%-- End Proof -- %%%%%%%%%%%%%%%%%%%%%%%%

%%%%%%%%%%**********%%%%%%%%%%**********%%%%%%%%%%**********%%%%%%%%%%
%%%%%%%%%%**********%%%%%%%%%%**********%%%%%%%%%%**********%%%%%%%%%%
\section{Numerical examples \vspace{-2mm}} \label{sec5} \setcounter{equation}{0} % Section 5

\ \ \ In this section, using two examples, we illustrate our theoretical results numerically by computing state feedback controllers which solve $\omega$-stabilization problems of interest. The first example considers a 1D heat equation with memory defined on the unit interval and the second example considers a 2D heat equation with memory defined on the unit square. \vspace{-3mm}

%%%%%%%%%%**********%%%%%%%%%%**********%%%%%%%%%%**********%%%%%%%%%%
%%%%%%%%%%**********%%%%%%%%%%**********%%%%%%%%%%**********%%%%%%%%%%
\subsection{\textbf{Example 1: 1D heat equation with memory}}

\ \ \ Consider the heat equation with memory \eqref{eq:HM_PIDE}-\eqref{eq:HM_BC} with $\Omega=(0,1)$, $\eta=0.01$, $\kappa=0.01$, $m=1$ and input shape function $b_1$ defined as follows: $b_1(\xi)=10$ if $\xi\in (0.1,0.8)$ and $b_1(\xi)=0$ otherwise. For these parameters we have
$\omega_0=\kappa+\frac{1}{\eta}=100.01$. The eigenvalues of the negative of the Laplacian operator $-\Delta: L^2(0,1)\to L^2(0,1)$ are $\lambda_j=j^2\pi^2$ for $j\in\nline$ and the corresponding eigenvectors are $\psi_j$ for $j\in\nline$, where
$\psi_j(\xi)=\sqrt{2}\sin(j\pi \xi)$ for all $\xi\in(0,1)$. These eigenvectors form an orthonormal basis for $L^2(0,1)$. Substituting for $\lambda_j$ in \eqref{eq:Aeigenvalues} we get that the eigenvalues of $A$ are \vspace{-1mm}
$$ \mu_j^{\pm} = \frac{-10^{-2} (1 + j^2\pi^2) \pm \sqrt{10^{-4} (1 + j^2 \pi^2)^2 - 4 j^2 \pi^2 (1 + 10^{-4})} }{2} \FORALL j\in \nline. \vspace{-1mm} $$
The first few eigenvalues of $A$ are $\mu_1^{ \pm}=-0.05\pm3.14i$, $\mu^{\pm}_2=-0.2\pm 6.28i$. Clearly, the open-loop response of \eqref{eq:HM_PIDE}-\eqref{eq:HM_BC} (for the parameters considered here) is slow. In this example, we want to construct a state feedback controller such that the closed-loop response has a decay rate of 1, i.e. we want to solve Problem \ref{prob:wstab} with $\omega=1<\omega_0$.

We will first verify that the pair $(A,B)$ is $\omega$-stabilizable for $\omega=1$. The unstable eigenvalues of $A+I$ are $\mu_1^{ \pm}+1=0.95\pm3.14i$, $\mu^{\pm}_2+1=0.80\pm 6.28i$, $\mu_3^{ \pm}+1=0.55 \pm 9.41i $ and $\mu_4^{ \pm}+1=0.21 \pm 12.54i$. So the set $\Jscr$ (defined in Theorem \ref{thm:ABstab}) is $\{1,2,3,4\}$. For any $\sbm{p\\q}\in X$, we have $B^*\sbm{p\\q}=\langle b_1, p\rangle_{L^2(0,1)}$, see \eqref{eq:Bdef-star}. A simple calculation gives \vspace{-2mm}
$$ B^{*}\bbm{\psi_1\\ 0} = 7.92, \quad B^{*}\bbm{\psi_2\\ 0} = 1.13, \quad B^{*}\bbm{\psi_3\\ 0} = 0.42, \quad
B^{*}\bbm{\psi_4\\ 0} = 1.26, \vspace{-2mm} $$
i.e. the condition \eqref{eq:Borth} holds. It now follows from Theorem \ref{thm:ABstab} that $(A,B)$ is $\omega$-stabilizable for $\omega=1$.

For each $n \geq 3$, consider the set of $n-1$ hat functions $\left\{\phi_1^n, \phi_2^n, \ldots \phi_{n-1}^{n}\right\}$ defined on $[0,1]$ as follows: \vspace{-2mm}
\begin{equation}\label{eq: basis functions}
 \phi^n_{j}(\xi)=\begin{cases}
   n\left(\xi-\frac{j-1}{n}\right) \quad &\forall \xi\in \left[\frac{j-1}{n},\frac{j}{n}\right),\\
   n\left(\frac{j+1}{n}-\xi\right) \quad &\forall \xi\in \left[\frac{j}{n},\frac{j+1}{n}\right],\\
   0 \quad &\forall \xi \notin \left[\frac{j-1}{n},\frac{j+1}{n}\right].
\end{cases} \vspace{-2mm}
\end{equation}
Clearly each $\phi^n_j\in H^1_0(0,1)$. Let $H_n$ be the span of the functions $\{\phi^n_1,\phi^n_2,\ldots\phi^n_{n-1}\}$ for $n\geq 3$ and take $H_1=H_2=H_3$. Define $V_n=H_n\times H_n$. From \cite[Section 3.2]{Ci:1978} we have that for each $q\in H^1_0(0,1)$ there exists a sequence $(q_n)_{n\in \nline}$ such that $q_n\in H_n$ for each $n\in\nline$ and \vspace{-1mm}
$$ \lim_{n\to \infty} \|q_n-q\|_{H^1_0(0,1)}=0. \vspace{-1mm}$$
From this it follows that the sequence of finite-dimensional subspaces $(V_n)_{n\in\nline}$ satisfy Assumption \ref{as:subspace}. We will use $(V_n)_{n\in\nline}$ to derive the finite-dimensional approximations of \eqref{eq:HM_absomega}.

We have now verified all the hypothesis in Theorem \ref{thm:main_result}. Suppose that $Q=2 I$ and $R=1$. From Theorem \ref{thm:main_result} it follows that there exists a unique nonnegative solution $\Pi\in\Lscr(X)$ to \eqref{eq:Riccati} such that $K_\infty=-R^{-1}B^{*}\Pi \in \Lscr(X,\rline)$ stabilizes \eqref{eq:HM_absomega} and also that there exists a unique nonnegative solution $\Pi_n\in\Lscr(V_n)$ to \eqref{eq:Riccati_approx} for each $n$ sufficiently large such that $K_n=-R^{-1}B^{*}_n\Pi_n\in\Lscr(V_n,\rline)$ stabilizes \eqref{eq:approx system}. Via the Riesz representation theorem it follows that there exist $\alpha\in L^2(0,1)$ and $\beta\in H^1_0(0,1)$ such that \vspace{-3mm}
$$ K_\infty \bbm{p \\ q}= \langle\alpha, p \rangle_{L^2(0,1)} + \langle \beta, q \rangle_{H^1_0(0,1)} \FORALL \bbm{p \\ q} \in X \vspace{-3mm}$$
and $\alpha_n\in L^2(0,1)$ and $\beta_n\in H^1_0(0,1)$ such that \vspace{-3mm}
$$ K_n P_n \bbm{p \\ q}= \langle\alpha_n, p \rangle_{L^2(0,1)} + \langle \beta_n, q \rangle_{H^1_0(0,1)} \FORALL \bbm{p \\ q} \in X. \vspace{-3mm} $$
For different values of $n$, we have solved \eqref{eq:Riccati_approx} numerically (by considering its equivalent matrix representation) to find $\Pi_n$, then computed $K_n$ using it and finally obtained $\alpha_n$ and $\beta_n$. From Table 1 it is evident that $\|\alpha_{n+1} - \alpha_n\|_{L^2(0,1)}$ and $\|\beta_{n+1} - \beta_n\|_{H^1_0(0,1)}$ become smaller as $n$ increases, which indicates that $\alpha_n$ and $\beta_n$ are converging to a limit (in $L^2(0,1)$ and $H^1_0(0,1)$, respectively) as $n$ tends to infinity. This is even more evident from the plots of $\alpha_n$ and $\frac{\dd \beta_n}{\dd \zeta}$ shown in Figure 2. All of these illustrate the convergence of the controllers gains claimed in \eqref{eq:Con-gain-conv}. \vspace{-5mm}

%%%%%%%%%%%%%%%%%%%%%%%%%%%%%%%%%%%%%%%%%%%%%%%%%%%%%%%%%%%%%%%%%%%%
%%%%%%%%%%%%%%%%%%%%%%%%%% Table 1 %%%%%%%%%%%%%%%%%%%%%%%%%%%%%%%%%
\begin{table}[ht]
 \centering
 \[
 \begin{array}{|c|c|c|}
   \hline
    n & \|\alpha_{n+1}-\alpha_{n}\|_{L^2(0,1)} & \|\beta_{n+1}-\beta_{n}\|_{H^1_0(0,1)} \\[0.5ex] \hline
    10  & 0.176 & 4.968 \\ \hline
    20  & 0.042 & 2.747 \\ \hline
    50  & 9.90 \times 10^{-3} & 0.825 \\ \hline
    100  & 1.60 \times 10^{-3} & 0.579 \\ \hline
    500 & 1.00 \times 10^{-4} & 0.086 \\ \hline
 %   1000 & 9.00 \times 10^{-6} & 0.012 \\ \hline
 \end{array}
 \]
\vspace{-5mm}

\parbox{5.2in}{\caption{\small{The values of \(\|\alpha_{n+1}-\alpha_{n}\|_{L^2(0,1)}\) and \(\|\beta_{n+1}-\beta_{n}\|_{H^1_0(0,1)}\) approach zero for large $n$, which indicates that $\alpha_n$ and $\beta_n$ are converging to a limit (in $L^2(0,1)$ and $H^1_0(0,1)$, respectively) as $n$ tends to infinity.\vspace{-2mm}}}}
\label{tab:convex1}
\end{table}
%%%%%%%%%%%%%%%%%%%%%%%%%%%%%%%%%%%%%%%%%%%%%%%%%%%%%%%%%%%%%%%%%%%%
%%%%%%%%%%%%%%%%%%%%%%%%% End Table %%%%%%%%%%%%%%%%%%%%%%%%%%%%%%%%

From the plots in Figure 2 it is evident that $\alpha_{50}$ and $\beta_{50}$ are close to the limit of the $\alpha_n$s and $\beta_n$s, respectively. So we choose the desired controller gain to be $K_\omega=K_{50}P_{50}$. When $n$ is sufficiently large, then \eqref{eq:approx system} is a good approximation of the infinite-dimensional dynamics \eqref{eq:HM_absomega}. For the purpose of this numerical example, we will suppose that \eqref{eq:approx system} with $n=1000$ is a good approximation of \eqref{eq:HM_absomega} and implement our feedback control law (with the chosen gain $K_{50}P_{50}$) on it. Figure 3 shows the eigenvalues of $A_{1000}+I$ and $A_{1000}+I+B_{1000}K_{50}P_{50}$. As expected, the latter eigenvalues are contained in the left half of the complex plane indicating that $K_\omega=K_{50}P_{50}$ solves the $\omega$-stabilization problem in this example, which supports the last statement of Theorem \ref{thm:main_result}. \vspace{-10mm}

%\m\vspace{-40mm}

$$\vspace{-25mm}$$

%%%%%%%%%%%%%%%%%%%%%%%%%%%%%%%%%%%%%%%%%%%%%%%%%%%%%%%%%%%%%%%%%%%%
%%%%%%%%%%%%%%%%%%%%%%%%%% Figure 2 %%%%%%%%%%%%%%%%%%%%%%%%%%%%%%%%
$$ \vspace{-4mm}
    % First subfigure
    {\includegraphics[width=0.48\textwidth]{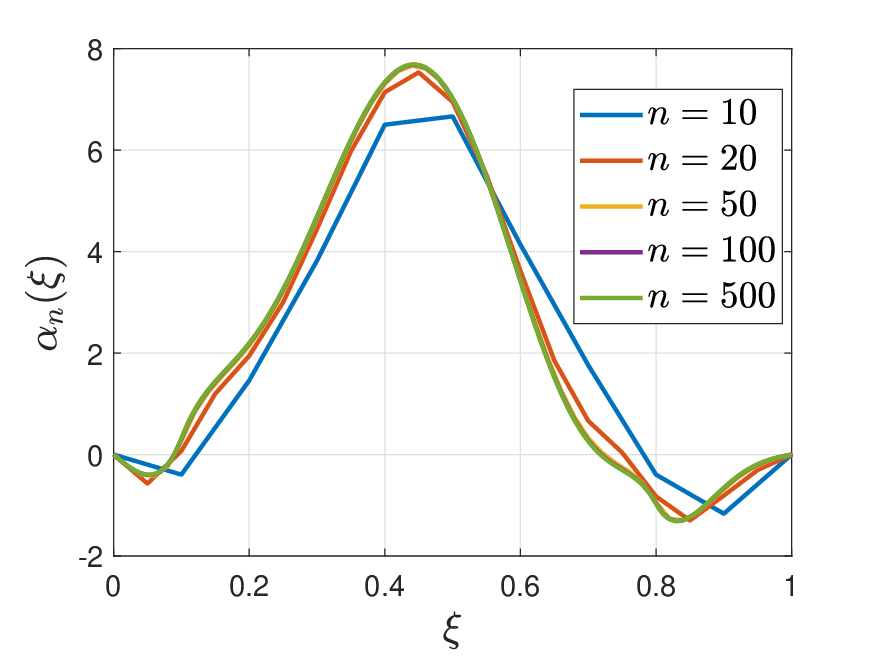}}
    % Second subfigure
    {\includegraphics[width=0.48\textwidth]{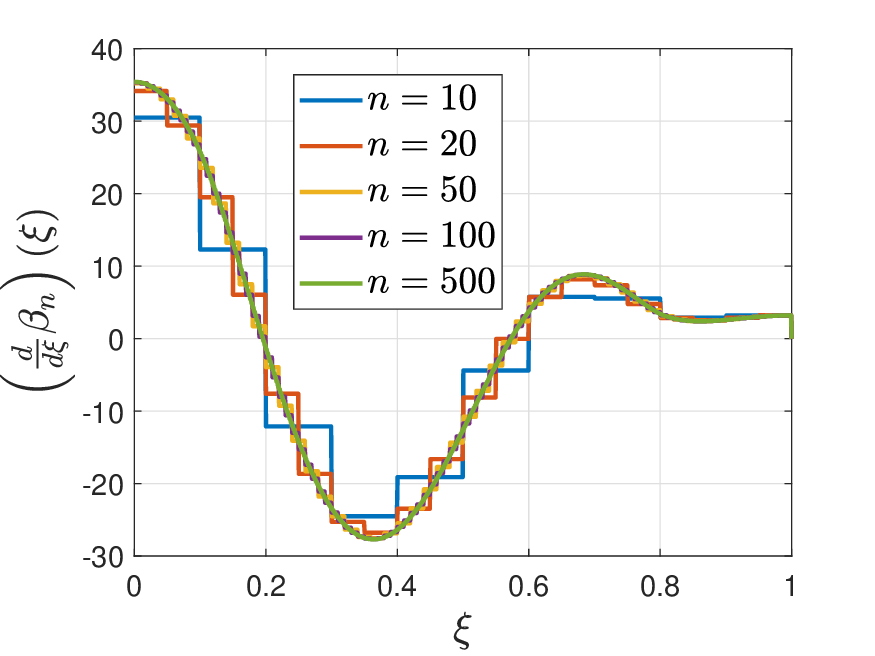}} $$
\begin{center}
{\parbox{5.2in}{\small{Figure 2. Plots of $\alpha_{n}$ and $\frac{\dd\beta_n}{\dd\xi}$ for different values of $n$ indicate that $\alpha_n$ and $\beta_n$ converge to a limit in $L^2(0,1)$ and $H^1_0(0,1)$, respectively, as $n\to\infty$.}}} \vspace{-4mm}
\end{center}
%%%%%%%%%%%%%%%%%%%%%%%%%%%%%%%%%%%%%%%%%%%%%%%%%%%%%%%%%%%%%%%%%%%%
%%%%%%%%%%%%%%%%%%%%%%%%%% End Figure %%%%%%%%%%%%%%%%%%%%%%%%%%%%%%

%%%%%%%%%%%%%%%%%%%%%%%%%%%%%%%%%%%%%%%%%%%%%%%%%%%%%%%%%%%%%%%%%%%%
%%%%%%%%%%%%%%%%%%%%%%%%%% Figure 3 %%%%%%%%%%%%%%%%%%%%%%%%%%%%%%%%
$$ \vspace{-2mm}
    % First subfigure
    {\includegraphics[width=0.48\textwidth]{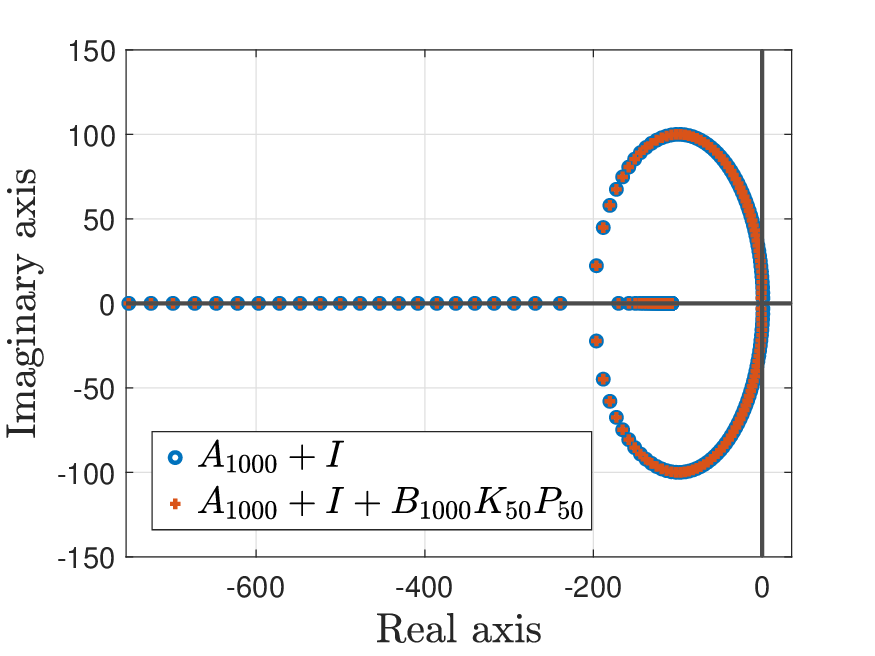}
    }
    % Second subfigure
    {\includegraphics[width=0.48\textwidth]{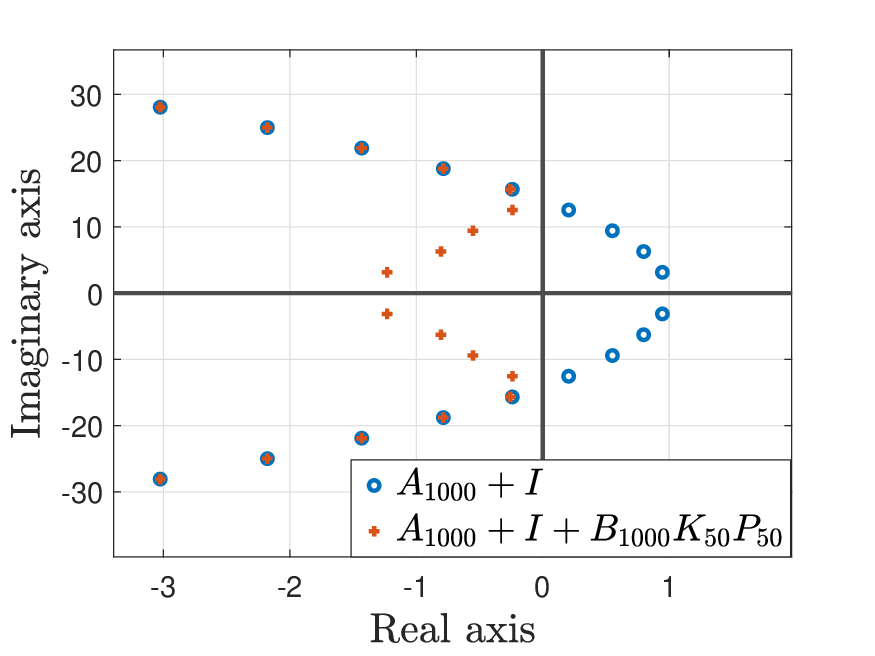}
    }$$
\begin{center}
    \parbox{5.2in}{{\small{Figure 3. The plot on the left shows the eigenvalues of $A_{1000} + I$ and $A_{1000}+I+B_{1000}K_{50}P_{50}$ and the plot on the right shows the zoomed-in view of the same eigenvalues near the origin. As predicted theoretically, 8 eigenvalues of $A_{1000} + I$ have a positive real part, but all the eigenvalues of $A_{1000}+I+B_{1000}K_{50}P_{50}$ have a negative real part. So $K_\omega=K_{50}P_{50}$ solves the $\omega$-stabilization problem for $\omega=1$. \vspace{-4mm}}}}
\end{center}
%    \label{fig:ex1eigen}
%\end{figure}
%%%%%%%%%%%%%%%%%%%%%%%%%%%%%%%%%%%%%%%%%%%%%%%%%%%%%%%%%%%%%%%%%%%%
%%%%%%%%%%%%%%%%%%%%%%%%%% End Figure %%%%%%%%%%%%%%%%%%%%%%%%%%%%%%

%%%%%%%%%%%%%%%%%%%%%%%%%%%%%%%%%%%%%%%%%%%%%%%%%%%%%%%%%%%%%%%%%%%%
%%%%%%%%%%%%%%%%%%%%%%%%%% Figure 4 %%%%%%%%%%%%%%%%%%%%%%%%%%%%%%%%
$$    % First subfigure
    {\includegraphics[width=0.48\textwidth]{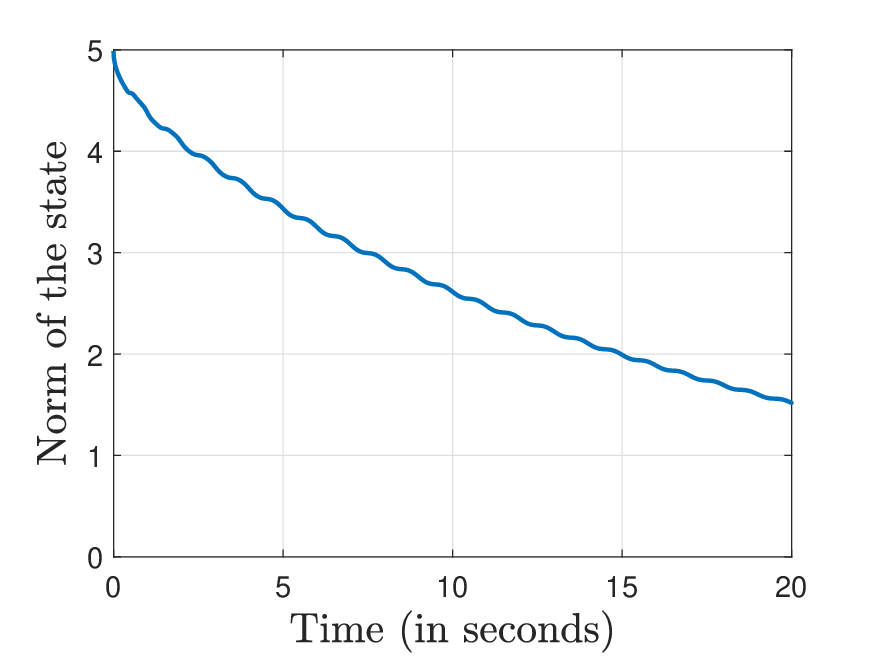}
    }
    % Second subfigure
    {\includegraphics[width=0.48\textwidth]{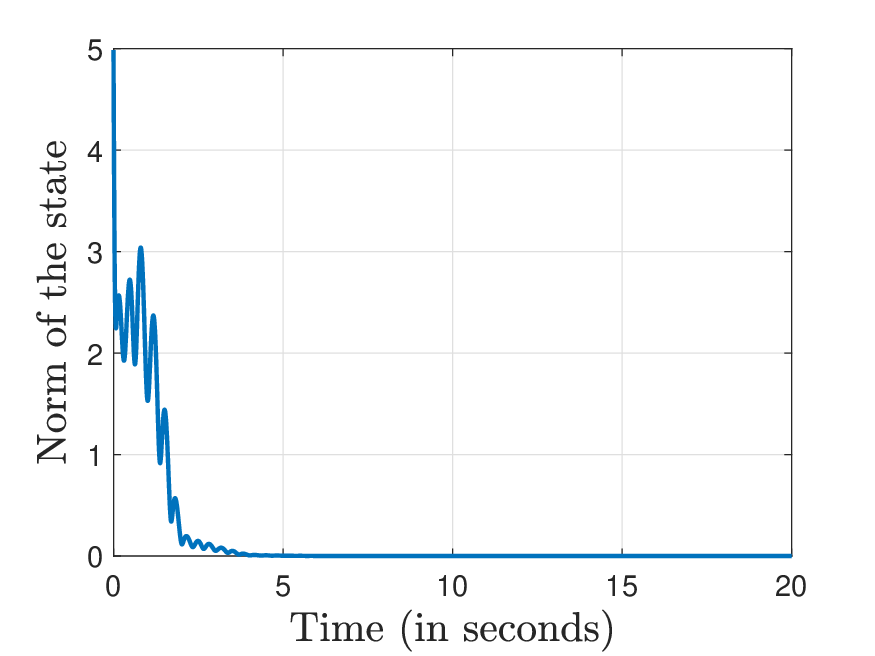}
    }$$
\begin{center}
\parbox{5.2in}{{\small{Figure 4. The plot on the left shows the open-loop response of \eqref{eq:HM_abs} to an initial state while the plot on the right shows the closed-loop response for the same initial state with the controller gain $K_{50}P_{50}$. As expected, the open-loop decay is sluggish, while the closed-loop decay is rapid.}}}
\end{center}
%%%%%%%%%%%%%%%%%%%%%%%%%%%%%%%%%%%%%%%%%%%%%%%%%%%%%%%%%%%%%%%%%%%%
%%%%%%%%%%%%%%%%%%%%%%%%%% End Figure %%%%%%%%%%%%%%%%%%%%%%%%%%%%%%

Figure 4 shows the response of \eqref{eq:HM_abs} to the initial state $\sbm{y^0 \\ z^0}=\sbm{-5\\0}$ when $u=0$ (i.e. the open-loop dynamics) and when $u=K_{50}P_{50}\sbm{y \\ z}$ (i.e. the closed-loop dynamics). The open-loop response decays slowly as expected, while the closed-loop response decays rapidly as desired.

%%%%%%%%%%**********%%%%%%%%%%**********%%%%%%%%%%**********%%%%%%%%%%
%%%%%%%%%%**********%%%%%%%%%%**********%%%%%%%%%%**********%%%%%%%%%%
\subsection{\textbf{Example 2: 2D heat equation with memory}}

\ \ \ Consider the heat equation with memory \eqref{eq:HM_PIDE}-\eqref{eq:HM_BC} with $\Omega=(0,1)\times(0,1)$, $\eta=0.005$, $\kappa=0.001$, $m=2$ and input shape functions $b_1,b_2$ defined as follows:
\begin{align*}
    b_1(\xi_1,\xi_2)=&\begin{cases}
        &5 \qquad\textrm{if}\quad \xi_1 \in (0.1,0.3) \ \ \textrm{and}\ \ \xi_2\in (0.1,0.5),\\
        &0 \qquad \textrm{otherwise},
    \end{cases}\\
    b_2(\xi_1,\xi_2)=&\begin{cases}
        &10 \qquad\textrm{if}\quad \xi_1 \in (0.5,0.7) \ \ \textrm{and}\ \ \xi_2\in (0.5,0.9), \\
        &0 \qquad \textrm{otherwise}.
    \end{cases}
\end{align*}
For these parameters we have $\omega_0=\kappa+ \frac{1}{\eta} =200.001$. The eigenvalues of the negative Laplacian operator $-\Delta: L^2(\Omega)\to L^2(\Omega)$ are $\lambda_{j,k}=(j^2+k^2)\pi^2$ for $j,k\in\nline$ and the corresponding eigenvectors are $\psi_{j,k}$ for $j,k\in\nline$, where $\psi_{j,k}(\xi_1,\xi_2)=2\sin(j\pi \xi_1)\sin(k\pi \xi_2)$ for all $\xi_1,\xi_2\in(0,1)$. These eigenvectors form an orthonormal basis for $L^2(\Omega)$. So the spectral properties of the negative Laplacian detailed in Theorem \ref{thm:Laplacian} hold for the negative Laplacian on the square (even though its boundary is not $C^2$) and therefore the results developed in this paper can be applied to the equation considered in this example. Using \eqref{eq:Aeigenvalues} we can compute the first few eigenvalues of $A$ to be $-0.05 \pm 4.44 i$ and $-0.12 \pm 7.02 i$. Clearly, the open-loop response of \eqref{eq:HM_PIDE}-\eqref{eq:HM_BC} (for the parameters considered here) is slow. In this example, we want to construct a state feedback controller such that the closed-loop response has a decay rate of 0.5, i.e. we want to solve Problem \ref{prob:wstab} with $\omega=0.5<\omega_0$.

We will first verify that the pair $(A,B)$ is $\omega$-stabilizable for $\omega=0.5$. The eigenvalues of $A$ lying in $\overline{\cline^+_{-0.5}}$, the eigenvalues of $-\Delta$ which when substituted in \eqref{eq:Aeigenvalues} give these eigenvalues of $A$ and a basis for the eigenspaces of $-\Delta$ corresponding to these eigenvalues are shown in Table 2.
%%%%%%%%%%%%%%%%%%%%%%%%%%%%%%%%%%%%%%%%%%%%%%%%%%%%%%%%%%%%%%%%%%%%
%%%%%%%%%%%%%%%%%%%%%%%%%%% Table 2 %%%%%%%%%%%%%%%%%%%%%%%%%%%%%%%%
\begin{table}[H]
\centering
\begin{tabular}{|c|c|c|}
\hline
Eigenvalues of $A$ in ${\overline{\cline^+_{-0.5}}}^{}$ & Eigenvalues of $-\Delta$ & Basis for eigenspaces of $-\Delta$ \\
\hline
$-0.05 \pm 4.443i$ & $2\pi^2$ & $\psi_{1,1}$ \\
\hline
$-0.124 \pm 7.024i$ & $5\pi^2$ & $\psi_{1,2}, \psi_{2,1}$ \\
\hline
$-0.247 \pm 9.932i$ & $10\pi^2$ & $\psi_{1,3}, \psi_{3,1}$ \\
\hline
$-0.42 \pm 12.946i$ & $17\pi^2$ & $\psi_{1,4}, \psi_{4,1}$ \\
\hline
$-0.198 \pm 8.884i$ & $8\pi^2$ & $\psi_{2,2}$ \\
\hline
$-0.321 \pm 11.323i$ & $13\pi^2$ & $\psi_{2,3}, \psi_{3,2}$ \\
\hline
$-0.494 \pm 14.041i$ & $20\pi^2$ & $\psi_{2,4}, \psi_{4,2}$ \\
\hline
$-0.445 \pm 13.321i$ & $18\pi^2$ & $\psi_{3,3}$ \\
\hline
\end{tabular}
\parbox{5.2in}{\caption{\small{Eigenvalues of $A$ lying in $\overline{\cline^+_{-0.5}}$, the associated eigenvalues of $-\Delta$ and a basis for the eigenspaces of $-\Delta$ corresponding to these eigenvalues.}}}
\label{tab:EigenALapEignSp}
\end{table}
%%%%%%%%%%%%%%%%%%%%%%%%%%%%%%%%%%%%%%%%%%%%%%%%%%%%%%%%%%%%%%%%%%%%
%%%%%%%%%%%%%%%%%%%%%%%%%%% Table 2 %%%%%%%%%%%%%%%%%%%%%%%%%%%%%%%%
\vspace{-6mm}

We have
$$
B^{*}\bbm{p\\q}= \bbm{\langle b_1,p \rangle_{L^2(\Omega)} \\ \langle b_2,p \rangle_{L^2(\Omega)}}  \FORALL \bbm{p\\q}\in X, \vspace{-1mm}
$$
see \eqref{eq:Bdef-star}. Table 3 lists all the matrices that must be considered, according to Theorem \ref{thm:ABstab} and Remark \ref{rm:checking}, for verifying the $\omega$-stabilizability of the pair $(A,B)$. These matrices clearly have the desired rank and so the pair $(A,B)$ is $\omega$-stabilizable for $\omega=0.5$. \vspace{-1mm}

%%%%%%%%%%%%%%%%%%%%%%%%%%%%%%%%%%%%%%%%%%%%%%%%%%%%%%%%%%%%%%%%%%%%
%%%%%%%%%%%%%%%%%%%%%%%%%%% Table 3 %%%%%%%%%%%%%%%%%%%%%%%%%%%%%%%%
\begin{table}[H]
\centering
\begin{tabular}{|c|c|}
\hline
Matrix & Rank \\
\hline\rule{0pt}{4ex}
$B^{*}\bbm{\psi_{1,1}\\0}$=
$\bbm{ 0.35 \\ 1.13}$ &
1 \\[0.35cm]
\hline\rule{0pt}{4ex}
$\bbm{B^{*} \bbm{\psi_{1,2}\\ 0} & B^{*}\bbm{\psi_{2,1}\\ 0}}$=
$\bbm{0.33 & 0.54 \\ -1.08 & -0.67}$ &
2 \\[0.35cm]
\hline\rule{0pt}{4ex}
$\bbm{B^{*} \bbm{\psi_{1,3}\\ 0} & B^{*}\bbm{\psi_{3,1}\\ 0}}$ =
$\bbm{0.07 & 0.49 \\ 0.23 & -0.61}$ &
2 \\[0.35cm]
\hline\rule{0pt}{4ex}
$\bbm{B^{*} \bbm{\psi_{1,4}\\ 0} & B^{*}\bbm{\psi_{4,1}\\ 0}}$ =
$\bbm{-0.06 & 0.27 \\ 0.21 & 0.87}$ &
2 \\[0.35cm]
\hline\rule{0pt}{4ex}
$B^{*}\bbm{\psi_{2,2}\\0}$ =
$\bbm{0.51 \\ 0.63}$ &
1 \\[0.35cm]
\hline\rule{0pt}{4ex}
$\bbm{B^{*} \bbm{\psi_{2,3}\\ 0} & B^{*}\bbm{\psi_{3,2}\\ 0}}$ =
$\bbm{0.11 & 0.47 \\ -0.14 & 0.58}$ &
2 \\[0.35cm]
\hline\rule{0pt}{4ex}
$\bbm{B^{*} \bbm{\psi_{2,4}\\ 0} & B^{*}\bbm{\psi_{4,2}\\ 0}}$ =
$\bbm{-0.10 & 0.26 \\ -0.12 & -0.83}$ &
2 \\[0.35cm]
\hline\rule{0pt}{4ex}
$B^{*}\bbm{\psi_{3,3}\\0}$ =
$\bbm{0.10 \\ -0.13}$ &
1 \\[0.35cm]
\hline
\end{tabular}
\parbox{5.2in}{\caption{\small{Matrices that must be considered for verifying the $\omega$-stabilizability of $(A,B)$ and their ranks \vspace{-3mm}}}}
\label{tab:verifyex2}
\end{table}
%%%%%%%%%%%%%%%%%%%%%%%%%%%%%%%%%%%%%%%%%%%%%%%%%%%%%%%%%%%%%%%%%%%%
%%%%%%%%%%%%%%%%%%%%%%%%%%% End Table 3 %%%%%%%%%%%%%%%%%%%%%%%%%%%%

Let $H_n=\text{span}\{\psi_{j,k} \ | \ j,k=1,2,\ldots n \}$. Define $V_n=H_n\times H_n$. From \cite[Remark 3.6.3]{Ke:2015} we have that for each $q\in H^1_0(0,1)$ there exists a sequence $(q_n)_{n\in \nline}$ such that $q_n\in H_n$ for each $n\in\nline$ and \vspace{-2mm}
$$ \lim_{n\to \infty} \|q_n-q\|_{H^1_0(\Omega)}=0. \vspace{-2mm} $$
From this it follows that the finite-dimensional subspaces $(V_n)_{n\in\nline}$ satisfy Assumption \ref{as:subspace}. We will use $(V_n)_{n\in\nline}$ to derive the finite-dimensional approximations of \eqref{eq:HM_absomega}.

We have now verified all the hypothesis in Theorem \ref{thm:main_result}. Suppose that $Q=I$ and $R=\sbm{1& 0 \\ 0 & 1}$. From Theorem \ref{thm:main_result} it follows that there exists a unique nonnegative solution $\Pi\in\Lscr(X)$ to \eqref{eq:Riccati} such that $K_\infty=-R^{-1}B^{*}\Pi \in \Lscr(X,\rline^2)$ stabilizes \eqref{eq:HM_absomega} and also that there exists a unique nonnegative solution $\Pi_n\in\Lscr(V_n)$ to \eqref{eq:Riccati_approx} for each $n$ sufficiently large such that $K_n=-R^{-1}B^{*}_n\Pi_n\in\Lscr(V_n,\rline^2)$ stabilizes \eqref{eq:approx system}. Via the Riesz representation theorem it follows that there exist $\alpha_1,\alpha_2\in L^2(\Omega)$ and $\beta_1,\beta_2\in H^1_0(\Omega)$ such that \vspace{-1mm}

$$ K_\infty \bbm{p \\ q}= \bbm{ \langle\alpha_1, p \rangle_{L^2(\Omega)} + \langle \beta_1, q \rangle_{H^1_0(\Omega)} \\ \langle\alpha_2, p \rangle_{L^2(\Omega)} + \langle \beta_2, q \rangle_{H^1_0(\Omega)}  } \FORALL \bbm{p \\ q} \in X \vspace{-1mm} $$
and $\alpha_{1,n},\alpha_{2,n}\in L^2(\Omega)$ and $\beta_{1,n},\beta_{2,n}\in H^1_0(\Omega)$ such that \vspace{-2mm}
$$ K_n P_n \bbm{p \\ q}= \bbm{\langle\alpha_{1,n}, p \rangle_{L^2(\Omega)} + \langle \beta_{1,n}, q \rangle_{H^1_0(\Omega)}\\ \langle\alpha_{2,n}, p \rangle_{L^2(\Omega)} + \langle \beta_{2,n}, q \rangle_{H^1_0(\Omega)}} \FORALL \bbm{p \\ q} \in X. \vspace{-2mm}$$
For different values of $n$, we have solved \eqref{eq:Riccati_approx} numerically (by considering its equivalent matrix representation) to find $\Pi_n$, then computed $K_n$ using it and finally obtained $\alpha_{1,n}$,$\alpha_{2,n}$ and $\beta_{1,n},\beta_{2,n}$. From Tables 4 and 5 it is evident that $\|\alpha_{1,n+1} - \alpha_{1,n}\|_{L^2(\Omega)}$, $\|\alpha_{2,n+1} - \alpha_{2,n}\|_{L^2(\Omega)}$, $\|\beta_{1,n+1} - \beta_{1,n}\|_{H^1_0(\Omega)}$ and $\|\beta_{2,n+1} - \beta_{2,n}\|_{H^1_0(\Omega)}$ become smaller as $n$ increases. These indicate that as $n$ tends to infinity $\alpha_{1,n}$, $\alpha_{2,n}$ converge to certain limits in $L^2(\Omega)$ and $\beta_{1,n}$, $\beta_{2,n}$ converge to certain limits in $H^1_0(\Omega)$, illustrating the convergence of the controller gains claimed in \eqref{eq:Con-gain-conv}. Figures 5 and 6 show the plots of $\alpha_{1,10}, \beta_{1,10}$, $\alpha_{2,10}$ and $\beta_{2,10}$ which approximate $\alpha_1, \beta_1, \alpha_2$ and $\beta_2$, \vspace{-4mm} respectively.

%%%%%%%%%%%%%%%%%%%%%%%%%%%%%%%%%%%%%%%%%%%%%%%%%%%%%%%%%%%%%%%%%%%%
%%%%%%%%%%%%%%%%%%%%%%%%%%% Table 4 %%%%%%%%%%%%%%%%%%%%%%%%%%%%%%%%
\begin{table}[ht]
\centering
\[
\begin{array}{|c|c|c|}
\hline
n & \|\alpha_{1,n+1}-\alpha_{1,n}\|_{L^2(\Omega)} &  \|\beta_{1,n+1}-\beta_{1,n}\|_{H^1_0(\Omega)}  \\[0.3ex]
\hline
2  & 5.6769 &  5.7450 \\
\hline
5  & 0.2439 &  0.1450 \\
\hline
10  & 0.0169 &  0.0052 \\
\hline
15  & 0.0086 &  0.0024 \\
\hline
20 & 0.0023 &  0.0006 \\
\hline
\end{array}
\]

\vspace{-4mm}

\parbox{5.2in}{\caption{\small{The values of \(\|\alpha_{1,n+1}-\alpha_{1,n}\|_{L^2(0,1)}\) and \(\|\beta_{1,n+1}-\beta_{1,n}\|_{H^1_0(0,1)}\) approach zero for large $n$, which indicates that $\alpha_{1,n}$ and $\beta_{1,n}$ are converging to a limit (in $L^2(0,1)$ and $H^1_0(0,1)$, respectively) as $n$ tends to infinity. \vspace{-6mm}
}}}
\label{tab:convex2one}
\end{table}
%%%%%%%%%%%%%%%%%%%%%%%%%%%%%%%%%%%%%%%%%%%%%%%%%%%%%%%%%%%%%%%%%%%%
%%%%%%%%%%%%%%%%%%%%%%%%%%% End Table 4 %%%%%%%%%%%%%%%%%%%%%%%%%%%

%%%%%%%%%%%%%%%%%%%%%%%%%%%%%%%%%%%%%%%%%%%%%%%%%%%%%%%%%%%%%%%%%%%%
%%%%%%%%%%%%%%%%%%%%%%%%%%% Table 5 %%%%%%%%%%%%%%%%%%%%%%%%%%%%%%%%
\begin{table}[ht]
\centering
\[
\begin{array}{|c|c|c|}
\hline
n & \|\alpha_{2,n+1}-\alpha_{2,n}\|_{L^2(\Omega)} &  \|\beta_{2,n+1}-\beta_{2,n}\|_{H^1_0(\Omega)}  \\[0.3ex]
\hline
2  & 3.3007 &  6.9288 \\
\hline
5  & 0.5154 &  0.5358 \\
\hline
10  & 0.0375 &  0.0177 \\
\hline
15  & 0.0257 &  0.0096 \\
\hline
20 & 0.0053 &  0.0018 \\
\hline
\end{array}
\]

\vspace{-4mm}

\parbox{5.2in}{\caption{\small{The values of $\|\alpha_{2,n+1}-\alpha_{2,n}\|_{L^2(0,1)}$ and $\|\beta_{2,n+1}-\beta_{2,n}\|_{H^1_0(0,1)}$ approach zero for large $n$, which indicates that $\alpha_{2,n}$ and $\beta_{2,n}$ are converging to a limit (in $L^2(0,1)$ and $H^1_0(0,1)$, respectively) as $n$ tends to infinity. \vspace{-2mm}
}}}
\label{tab:convex2two}
\end{table}
%%%%%%%%%%%%%%%%%%%%%%%%%%%%%%%%%%%%%%%%%%%%%%%%%%%%%%%%%%%%%%%%%%%%
%%%%%%%%%%%%%%%%%%%%%%%%%%% End Table 5 %%%%%%%%%%%%%%%%%%%%%%%%%%%%

From Tables 4 and 5 it appears that $\alpha_{1,10}$, $\alpha_{2,10}$, $\beta_{1,10}$ and $\beta_{2,10}$ are close to the limits (as $n$ tends to infinity) of $\alpha_{1,n}$, $\alpha_{2,n}$, $\beta_{1,n}$ and $\beta_{2,n}$, respectively. Indeed, we have $\|\alpha_{1,25}-\alpha_{1,10}\|_{L^2(\Omega)}=0.045$, $\|\beta_{1,25}-\beta_{1,10}\|_{H_0^1(\Omega)}=0.013$, $ \| \alpha_{2,25}-$ $\alpha_{2,10} \|_{L^2(\Omega)}=0.083$ and $\|\beta_{2,25}-\beta_{2,10}\|_{H_0^1(\Omega)}=0.035$. On the basis of this, we choose the desired controller gain to be $K_\omega=K_{10}P_{10}$. When $n$ is sufficiently large, then \eqref{eq:approx system} is a good approximation of the infinite-dimensional dynamics \eqref{eq:HM_absomega}. For the purpose of this numerical example, we will suppose that \eqref{eq:approx system} with $n=25$ is a good approximation of \eqref{eq:HM_absomega} and implement our feedback control law (with the chosen gain $K_{10}P_{10}$) on it. Figure 7 shows the eigenvalues of $A_{25}+0.5I$ and $A_{25}+0.5I+B_{25}K_{10}P_{10}$. As expected, the latter eigenvalues are contained in the left half of the complex plane indicating that $K_\omega=K_{10}P_{10}$ solves the $\omega$-stabilization problem in this example, which supports the last statement of Theorem \ref{thm:main_result}. \vspace{-5mm}

%%%%%%%%%%%%%%%%%%%%%%%%%%%%%%%%%%%%%%%%%%%%%%%%%%%%%%%%%%%%%%%%%%%%
%%%%%%%%%%%%%%%%%%%%%%%%%% Figure 5 %%%%%%%%%%%%%%%%%%%%%%%%%%%%%%%%
$$
    % First subfigure
    {\includegraphics[width=0.48\textwidth]{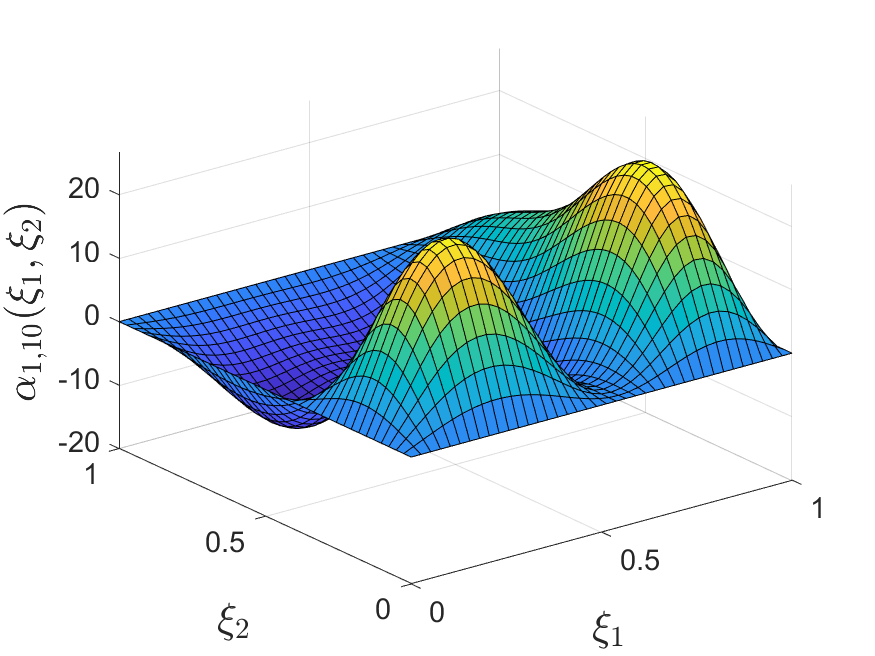}
    }
    % Second subfigure
    {\includegraphics[width=0.48\textwidth]{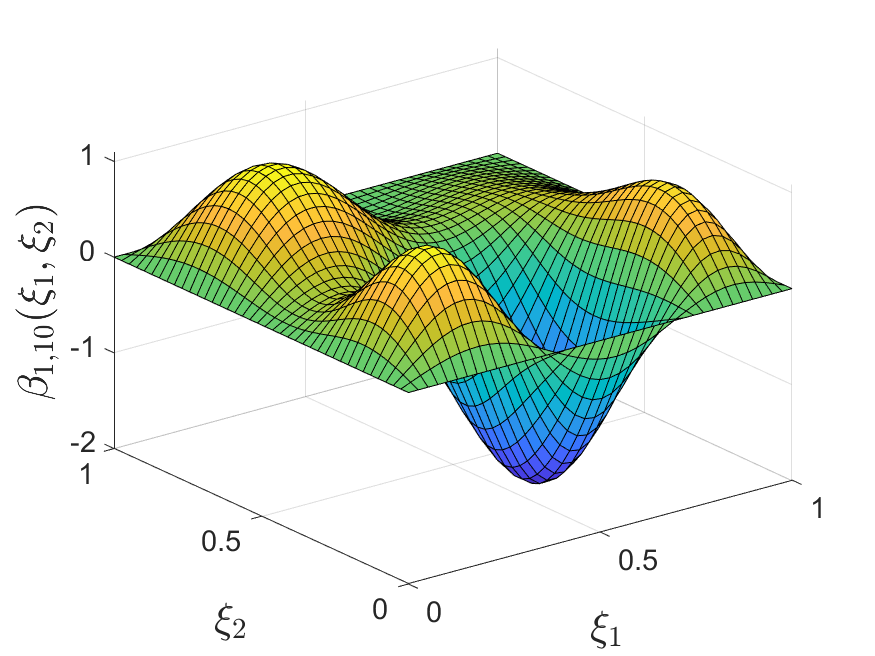}
    }$$
\begin{center}
 \parbox{5.2in}{{\small{Figure 5. Plots of $\alpha_{1,10}$ and $\beta_{1,10}$. These functions approximate the functions $\alpha_1$ and $\beta_1$.\vspace{-5mm}}}}
\end{center}
%%%%%%%%%%%%%%%%%%%%%%%%%%%%%%%%%%%%%%%%%%%%%%%%%%%%%%%%%%%%%%%%%%%%
%%%%%%%%%%%%%%%%%%%%%%%%%% End Figure %%%%%%%%%%%%%%%%%%%%%%%%%%%%%%

%%%%%%%%%%%%%%%%%%%%%%%%%%%%%%%%%%%%%%%%%%%%%%%%%%%%%%%%%%%%%%%%%%%%
%%%%%%%%%%%%%%%%%%%%%%%%%% Figure 6 %%%%%%%%%%%%%%%%%%%%%%%%%%%%%%%%
$$
    % First subfigure
    {\includegraphics[width=0.48\textwidth]{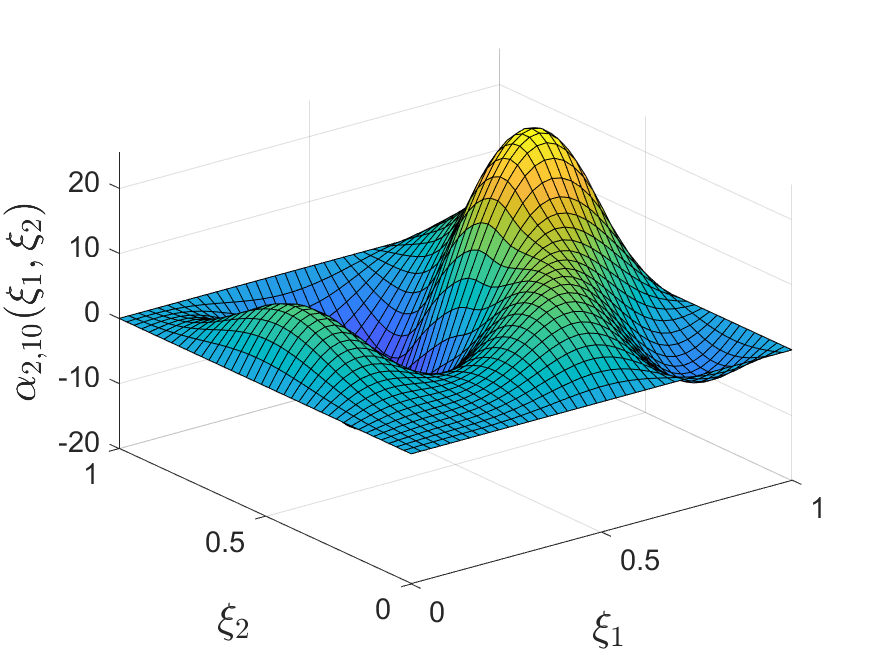}
    }
    % Second subfigure
    {\includegraphics[width=0.48\textwidth]{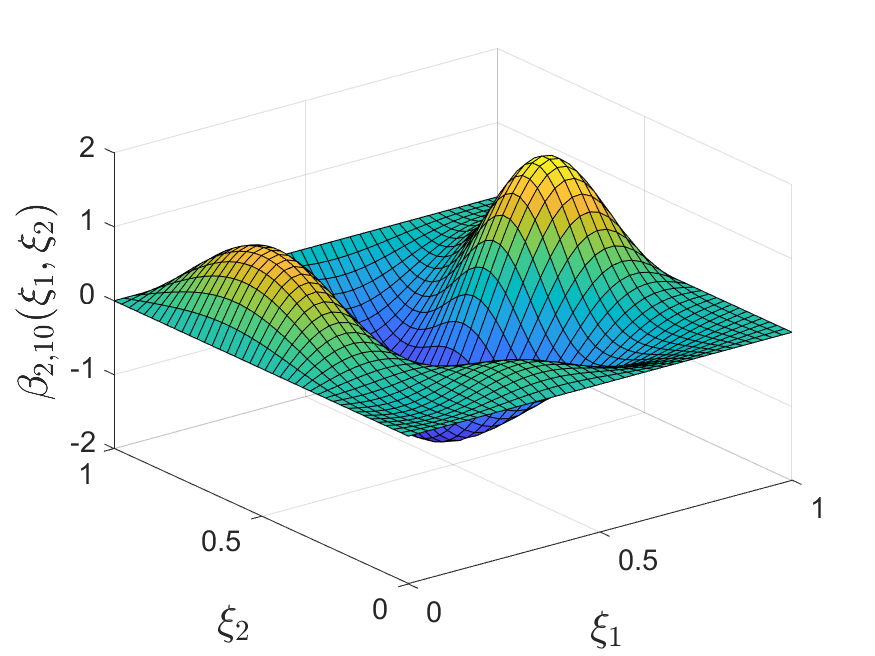}
    }$$
\begin{center}
\parbox{5.2in}{{\small{Figure 6. Plots of $\alpha_{2,10}$ and $\beta_{2,10}$. These functions approximate the functions $\alpha_2$ and $\beta_2$.\vspace{3mm}}}}
\end{center}
%%%%%%%%%%%%%%%%%%%%%%%%%%%%%%%%%%%%%%%%%%%%%%%%%%%%%%%%%%%%%%%%%%%%
%%%%%%%%%%%%%%%%%%%%%%%%%% End Figure %%%%%%%%%%%%%%%%%%%%%%%%%%%%%%

Figure 8 shows the response of \eqref{eq:HM_abs} to the initial state $\sbm{y^0 \\ z^0}=\sbm{2\\0}$ when $u=0$ (i.e. the open-loop dynamics) and when $u=K_{10}P_{10}\sbm{y \\ z}$ (i.e. the closed-loop dynamics). The open-loop response decays slowly as expected, while the closed-loop response decays rapidly as desired.

%%%%%%%%%%%%%%%%%%%%%%%%%%%%%%%%%%%%%%%%%%%%%%%%%%%%%%%%%%%%%%%%%%%%
%%%%%%%%%%%%%%%%%%%%%%%%%% Figure 7 %%%%%%%%%%%%%%%%%%%%%%%%%%%%%%%%
$$  % First subfigure
    {\includegraphics[width=0.47\textwidth]{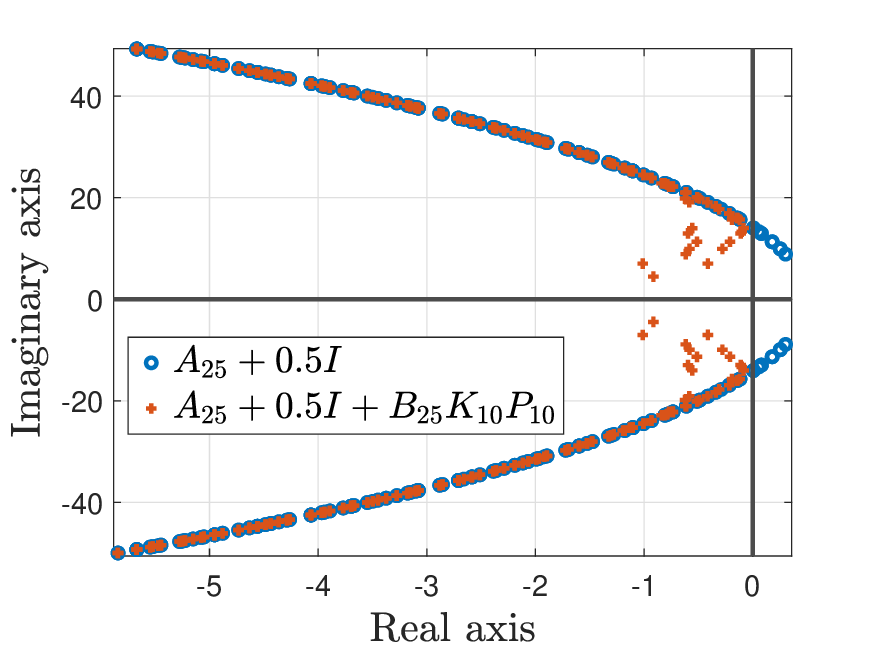}
    }
    % Second subfigure
    {\includegraphics[width=0.47\textwidth]{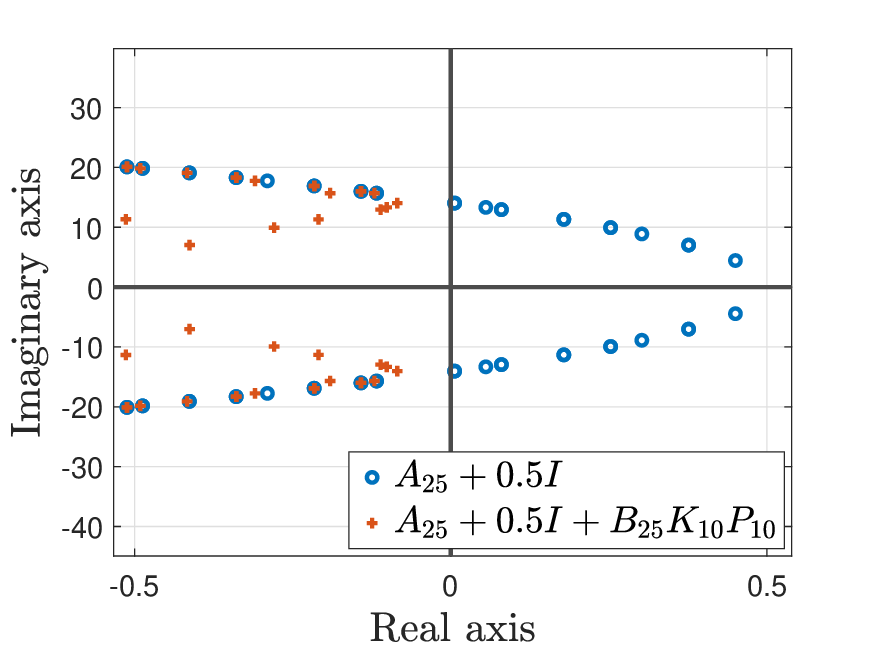}
    }$$
\begin{center}
 \parbox{5.2in}{{\small{Figure 7. The plot on the left shows the eigenvalues of $A_{25} + 0.5I$ and $A_{25}+0.5I+B_{25}K_{10}P_{10}$ and the plot on the right shows the zoomed-in view of the same eigenvalues near the origin. As predicted theoretically, 16 eigenvalues of $A_{25} + 0.5I$ have a positive real part, but all the eigenvalues of $A_{25}+0.5I+B_{25}K_{10}P_{10}$ have a negative real part. Hence $K_\omega=K_{10}P_{10}$ solves the $\omega$-stabilization problem for $\omega=0.5$. \vspace{-4mm}}}}
\end{center}
%%%%%%%%%%%%%%%%%%%%%%%%%%%%%%%%%%%%%%%%%%%%%%%%%%%%%%%%%%%%%%%%%%%%
%%%%%%%%%%%%%%%%%%%%%%%%%% End Figure %%%%%%%%%%%%%%%%%%%%%%%%%%%%%%

%%%%%%%%%%%%%%%%%%%%%%%%%%%%%%%%%%%%%%%%%%%%%%%%%%%%%%%%%%%%%%%%%%%%
%%%%%%%%%%%%%%%%%%%%%%%%%% Figure 8 %%%%%%%%%%%%%%%%%%%%%%%%%%%%%%%%
$$  % First subfigure
    {\includegraphics[width=0.47\textwidth]{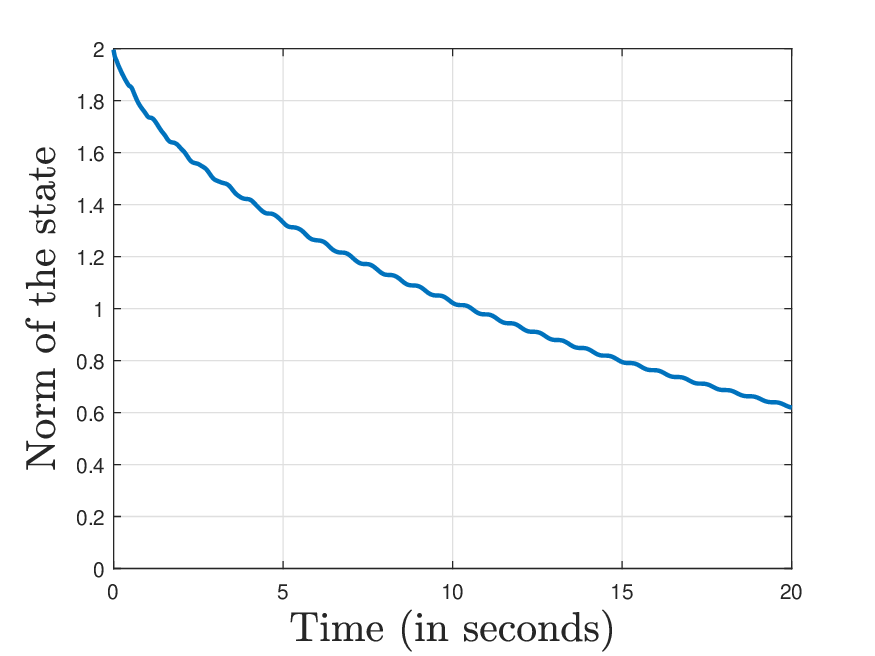}
    }
    % Second subfigure
    {\includegraphics[width=0.47\textwidth]{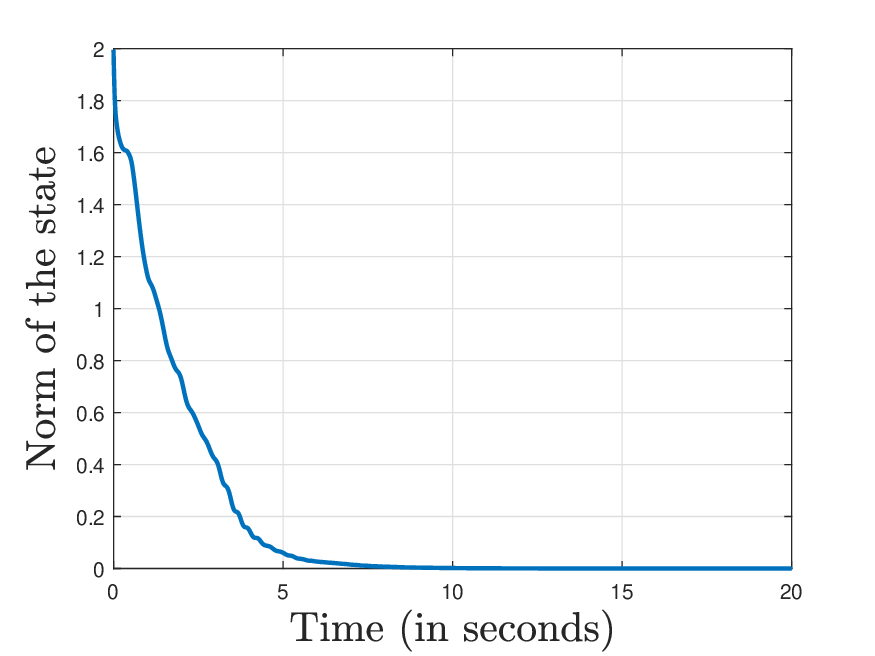}
    }$$
\begin{center}
\parbox{5.2in}{{\small{Figure 8. The plot on the left shows the open-loop response of \eqref{eq:HM_abs} to an initial state while the plot on the right shows the closed-loop response for the same initial state with the controller gain $K_{10}P_{10}$. As expected, the open-loop decay is sluggish, while the closed-loop decay is rapid.}}}
\end{center}
%%%%%%%%%%%%%%%%%%%%%%%%%%%%%%%%%%%%%%%%%%%%%%%%%%%%%%%%%%%%%%%%%%%%
%%%%%%%%%%%%%%%%%%%%%%%%%% End Figure %%%%%%%%%%%%%%%%%%%%%%%%%%%%%%

%%%%%%%%%%**********%%%%%%%%%%**********%%%%%%%%%%**********%%%%%%%%%%
%%%%%%%%%%**********%%%%%%%%%%**********%%%%%%%%%%**********%%%%%%%%%%
\section{Conclusion} \label{sec6} \vspace{-1mm}
\setcounter{equation}{0} % Section 6

\ \ \ In this work, for some $\omega>0$, we have addressed the $\omega$-stabilization problem for a heat equation with memory which is defined on a bounded domain in $\rline^d$ and is driven by $m$ control inputs acting on the interior of the domain. The state and control operators associated with the heat equation with memory are denoted by $A$ and $B$, respectively. Under the verifiable assumption that the pair $(A,B)$ is $\omega$-stabilizable, we have presented a numerical scheme for constructing a state-feedback controller gain $K_\omega$ such that $A+\omega I +B K_\omega$ is exponentially stable. Our scheme involves constructing a sequence of finite-dimensional approximations $(A_n)_{n\in\nline}$ and $(B_n)_{n\in\nline}$ of $A$ and $B$, respectively, and using them to construct approximate solutions to an appropriate Riccati equation associated with the heat equation with memory, and then using these solutions to obtain the required controller gain $K_\omega$. We have proved the validity of our scheme by establishing the crucial uniform (in $n$) stabilizability of the pair $(A_n+\omega I, B_n)$ and then using the results from \cite{BaKu:1984}.

Suppose that the state of the heat equation with memory is not available and instead only a finite-dimensional output of the equation can be measured. In this case, the state-feedback controller developed in this work cannot be implemented to solve the $\omega$-stabilization problem for the heat equation with memory. However, an output-feedback controller which can be implemented can be developed. Indeed, let $C\in\Lscr(X,\rline^p)$ be the output operator associated with the measurement and assume that the pair $(A+\omega I, C)$ is detectable. (Conditions analogous to those in Theorem \ref{thm:ABstab} can be derived for verifying this assumption.) Consider the sequence $(C_n)_{n\in\nline}$, where $C_n$ is the restriction of $C$ to $V_n$ (here $V_n$ is the same space used to define $A_n$ and $B_n$). Adapting our proof for the uniform (in $n$) stabilizability of the pair $(A_n+\omega I, B_n)$, see Propositions \ref{pr:approxeig} and \ref{pr:unifstab}, we can prove the uniform (in $n$) detectability of the pair $(A_n+\omega I, C_n)$. This uniform detectability, the uniform stabilizability of $(A_n+\omega I, B_n)$ and Condition (C2) established in the proof of Theorem \ref{thm:main_result} constitute sufficient conditions for applying \cite[Theorem 4.12]{Cur:03} to construct a robust reduced-order output-feedback controller which solves the $\omega$-stabilization problem for the heat equation with memory. In particular, the semigroup $\tline^{cl}$ associated with the closed-loop of the heat equation with memory and the reduced-order output-feedback controller constructed based on \cite[Theorem 4.12]{Cur:03} will satisfy \eqref{eq:probbnd} for some $M,\epsilon>0$. We hope to explore the construction of such reduced-order finite-dimensional controllers in a future work. \vspace{-3mm}

%%%%%%%%%%%%%%%%%%%%%%%%%%%%%%%%%%%%%%%%%%%%%%%%%%%%%%%%%%%%%%%%%%%%%%%%%
%%%%%%%%%%%%%%%%%%%%%%%%%%%%%%%%%%%%%%%%%%%%%%%%%%%%%%%%%%%%%%%%%%%%%%%%%

\end{document}